\documentclass[twocolumn]{aastex631}
\usepackage{xcolor}

\newcommand {\dg}{$^\circ$}
\newcommand {\kms}{$\rm{km\ s^{-1}}$}

\begin{document}

\title{JWST NIRSpec observations of Supernova 1987A -- from the inner ejecta to the reverse shock }

\correspondingauthor{J. Larsson}
\email{josla@kth.se}

\author[0000-0003-0065-2933]{J. Larsson}
\affiliation{Department of Physics, KTH Royal Institute of Technology, The Oskar Klein Centre, AlbaNova, SE-106 91 Stockholm, Sweden}

\author[0000-0001-8532-3594]{C. Fransson}
\affiliation{Department of Astronomy, Stockholm University, The Oskar Klein Centre, AlbaNova, SE-106 91 Stockholm, Sweden}

\author[0000-0001-9855-8261]{B. Sargent}
\affiliation{Space Telescope Science Institute, 3700 San Martin Drive, Baltimore, MD 21218, USA}
\affiliation{Center for Astrophysical Sciences, The William H. Miller III Department of Physics and Astronomy, Johns Hopkins University, Baltimore, MD 21218, USA}

\author[0000-0003-4870-5547]{O.\ C.\ Jones}
\affiliation{UK Astronomy Technology Centre, Royal Observatory, Blackford Hill, Edinburgh, EH9 3HJ, UK}

\author[0000-0002-3875-1171]{M.\ J.\ Barlow}
\affiliation{Department of Physics and Astronomy, University College London (UCL), Gower Street, London WC1E 6BT, UK}

\author{P. Bouchet}
\affiliation{Laboratoire AIM Paris-Saclay, CNRS, Universite Paris Diderot, F-91191 Gif-sur-Yvette, France}

\author[0000-0002-0522-3743]{M.\ Meixner}
\affil{Stratospheric Observatory for Infrared Astronomy, NASA Ames Research Center, Mail Stop 204-14, Moffett Field, CA 94035, USA}
\affil{Jet Propulsion Laboratory, California Institute of Technology, 4800 Oak Grove Dr., Pasadena, CA 91109, USA}

\author[0000-0002-5797-2439]{J. A. D. L. Blommaert}
\affiliation{Astronomy and Astrophysics Research Group, Department of Physics and Astrophysics, Vrije Universiteit Brussel, Pleinlaan 2, B-1050 Brussels, Belgium}

\author[0000-0001-6492-7719]{A. Coulais}
\affiliation{LERMA, Observatoire de Paris, Universit\'e PSL, Sorbonne Universi\'e, CNRS, Paris, France}
\affiliation{Astrophysics Department CEA-Saclay, France}

\author[0000-0003-2238-1572]{O.\ D.\ Fox}
\affiliation{Space Telescope Science Institute, 3700 San Martin Drive, Baltimore, MD, 21218, USA}

\author{R. Gastaud}
\affiliation{Laboratoire AIM Paris-Saclay, CEA-IRFU/SAp, CNRS, Universite Paris Diderot, F-91191 Gif-sur-Yvette, France}

\author[0000-0002-2041-2462]{A.\ Glasse}
\affiliation{UK Astronomy Technology Centre, Royal Observatory, Blackford Hill, Edinburgh, EH9 3HJ, UK}

\author[0000-0002-2667-1676]{N.\ Habel}
\affil{Stratospheric Observatory for Infrared Astronomy, NASA Ames Research Center, Mail Stop 204-14, Moffett Field, CA 94035, USA}
\affil{Jet Propulsion Laboratory, California Institute of Technology, 4800 Oak Grove Dr., Pasadena, CA 91109, USA}

\author[0000-0002-2954-8622]{A.\ S.\ Hirschauer}
\affil{Space Telescope Science Institute, 3700 San Martin Drive, Baltimore, MD 21218, USA}

\author[0000-0002-4571-2306]{J.\ Hjorth}
\affil{DARK, Niels Bohr Institute, University of Copenhagen, Jagtvej 128, 2200 Copenhagen, Denmark}

\author{J.\ Jaspers}
\affil{Dublin Institute for Advanced Studies, School of Cosmic Physics, Astronomy \& Astrophysics Section, 31 Fitzwilliam Place,\\ Dublin 2, Ireland.}
\affil{Department of Experimental Physics, Maynooth University-National University of Ireland Maynooth, Maynooth, Co Kildare, Ireland}

\author[0000-0001-6872-2358]{P.\ J.\ Kavanagh}
\affil{Department of Experimental Physics, Maynooth University-National University of Ireland Maynooth, Maynooth, Co Kildare, Ireland}
\affil{Dublin Institute for Advanced Studies, School of Cosmic Physics, Astronomy \& Astrophysics Section, 31 Fitzwilliam Place,\\ Dublin 2, Ireland.}

\author{O. Krause}
\affiliation{Max-Planck-Institut fuer Astronomie, Koenigstuhl 17, D-69117 Heidelberg, Germany}

\author[0000-0003-0778-0321]{R.\ M.\ Lau}
\affil{NSF’s NOIR Lab 950 N. Cherry Avenue, Tucson, AZ 85721, USA}

\author[0000-0003-4023-8657]{L.\ Lenki\'{c}}
\affil{Stratospheric Observatory for Infrared Astronomy, NASA Ames Research Center, Mail Stop 204-14, Moffett Field, CA 94035, USA}

\author[0000-0001-6576-6339]{O.\ Nayak}
\affil{Space Telescope Science Institute, 3700 San Martin Drive, Baltimore, MD 21218, USA}

\author[0000-0002-4410-5387]{A.\ Rest}
\affil{Space Telescope Science Institute, 3700 San Martin Drive, Baltimore, MD 21218, USA}
\affil{Department of Physics and Astronomy, Johns Hopkins University, 3400 North Charles Street, Baltimore, MD 21218, USA}

\author[0000-0001-7380-3144]{T.\ Temim}
\affil{Department of Astrophysical Sciences, Princeton University, Princeton, NJ 08544, USA}

\author{T.\ Tikkanen}
\affiliation{School of Physics \& Astronomy, Space Research Centre, University of Leicester, Space Park Leicester, 92 Corporation Road, Leicester LE4 5SP, UK}

\author[0000-0002-4000-4394]{R.\ Wesson}
\affiliation{School of Physics and Astronomy, Cardiff University, Queen’s Buildings, The Parade, Cardiff, CF24 3AA, UK}

\author{G.\ S.\ Wright}
\affiliation{UK Astronomy Technology Centre, Royal Observatory, Blackford Hill, Edinburgh, EH9 3HJ, UK}

\begin{abstract}
We present initial results from {\it JWST} NIRSpec integral field unit observations of the nearby Supernova (SN) 1987A. The observations provide the first spatially-resolved spectroscopy of the ejecta and equatorial ring (ER) over the 1--5~$\mu$m range. We construct 3D emissivity maps of the [\ion{Fe}{1}]~$1.443\ \mu$m line from the inner ejecta and the \ion{He}{1}~1.083~$\mu$m line from the reverse shock (RS), where the former probes the explosion geometry and the latter traces the structure of the circumstellar medium. We also present a model for the integrated spectrum of the ejecta. The [\ion{Fe}{1}] 3D map reveals a highly-asymmetric morphology resembling a broken dipole, dominated by two large clumps with  velocities of $\sim 2300$~\kms. We also find evidence that the Fe-rich inner ejecta have started to interact with the RS. The RS surface traced by the \ion{He}{1} line extends from just inside the ER to higher latitudes on both sides of the ER with a half-opening angle $\sim 45$\dg, forming a bubble-like structure. The spectral model for the ejecta allows us to identify the many emission lines, including numerous H$_2$ lines. We find that the H$_2$ is most likely excited by far-UV emission, while the metal lines ratios are consistent with a combination of collisional excitation and recombination in the low-temperature ejecta. We also find several high-ionization coronal lines from the ER, requiring a temperature $\ga 2 \times 10^6$ K.

\end{abstract}


\section{Introduction} 
\label{sec:intro}

The recently-launched {\it JWST} \citep{Gardner2006} is revolutionizing our understanding of the infrared (IR) emission from a wide range of astrophysical phenomena. One of the targets observed by {\it JWST} in its first year of operation is the iconic Supernova (SN) 1987A (see \citealt{McCray1993,McCray2016} for reviews).  Owing to its proximity in the Large Magellanic Cloud (LMC), astronomers have been able to follow the evolution of this SN across the entire electromagnetic spectrum as it evolves into a SN remnant (SNR). The first {\it JWST} observations of SN~1987A were carried out as part of guaranteed time observation (GTO) program 1232 (PI: G. Wright), using NIRSpec \citep{Jakobsen2022} as well as the MIRI medium-resolution spectrometer (MRS; \citealt{Wells2015}) and Imager \citep{Bouchet2015,Wright2023}. These observations provide unprecedented information about the IR emission from the system and make it possible to address a wide range of scientific questions regarding the ejecta, circumstellar medium (CSM), and compact object. In this paper, we present initial results concerning the near-IR (NIR) emission based on the NIRSpec observations. 

The progenitor of SN~1987A was a blue supergiant (BSG; \citealt{Walborn1987}), which is thought to have been produced as a result of a binary merger \citep[e.g.,][]{Hillebrandt1989,Podsiadlowski1990,Menon2017,Ono2020,Orlando2020,Utrobin2021}, which may also have created the triple-ring nebula of CSM \citep{Morris2007,Morris2009}. The nebula comprises an inner equatorial ring (ER) with radius $\sim 0\farcs{8}$, as well as two larger outer rings (ORs) located above and below its plane. The ER and ORs have inclinations in the range $38-45$\dg\ \citep{Tziamtzis2011}, which causes them to appear elliptical as projected on the sky. 

The shock interaction between the ejecta and the ER has produced bright multiwavelength emission since $\sim 5000$~days post-explosion \citep[e.g.,][]{McCray2016}, but the IR, optical, and soft X-ray emission is currently fading \citep{Fransson2015,Larsson2019b,Arendt2020,Alp2021,Maitra2022,Kangas2023}, signaling that the dense ER is being destroyed by the shocks and that the blast wave has passed through it \citep{Fransson2015}. At the same time, the dense inner ejecta have continued their free expansion inside the ER, revealing a highly asymmetric distribution in increasingly great detail over time. 

Observations of SN~1987A in the NIR range date back to 1987. The early observations revealed strong broad emission lines from a wide range of elements in the ejecta \citep[e.g.,][]{Meikle1993}, as well as the first detection of CO and SiO in a SN \citep{Catchpole1988,Spyromilio1988,Elias1988,Meikle1989,Roche1991}. Later, NIR observations with the integral field unit (IFU) SINFONI at the VLT showed H$_2$ emission from the ejecta \citep{Fransson2016}, confirming theoretical models for molecule formation \citep{Culhane1995}. The SINFONI observations also revealed a rich spectrum of broad atomic lines from the inner ejecta, as well as narrow lines from the shocked gas in the ER \citep{Kjaer2007,Kjaer2010}. 

The brightest lines from the ejecta have been reconstructed in 3D, making use of the linear relation between ejecta velocity and distance in the freely expanding ejecta \citep{Kjaer2010,Larsson2016,Larsson2019a}. The 3D distributions are highly asymmetric, which reflects the conditions at the time of explosion and hence the explosion mechanism \citep[e.g.,][]{Sandoval2021,Gabler2021}. 

The {\it JWST} NIRSpec observations greatly improve our knowledge of the NIR emission from SN~1987A. The observations were carried out in IFU mode and provide spatially-resolved spectroscopy over the full  $1.0-5.2~\mu$m wavelength range. In this first paper on these observations, we present the integrated spectra of the ejecta and ER, and provide a spectral model for the ejecta. In addition, we take advantage of the IFU to create 3D maps of the [\ion{Fe}{1}]~$1.443~\mu \rm{m}$ emission from the inner ejecta and the \ion{He}{1}~$1.083~\mu \rm{m}$ emission from the reverse shock (RS). To the best of our knowledge, the most recent previous observations of these two lines were obtained in 1995 and 1992, respectively \citep{Fassia2002}.

The [\ion{Fe}{1}]~$1.443~\mu \rm{m}$ line is bright and not blended with any other strong lines, making it an excellent probe of the ejecta geometry and properties of the explosion. Importantly, it provides the first 3D view of Fe in the ejecta, which adds valuable new information compared to previous 3D maps of the  $1.65\ \mu$m line, which  is a blend of [\ion{Fe}{2}] and [\ion{Si}{1}] \citep{Kjaer2010,Larsson2016}. 

The \ion{He}{1}~$1.083~\mu \rm{m}$ line is by far the brightest line emitted from the RS in the NIR. The 3D emissivity of this line traces the CSM, which gives insight into the nature of the progenitor and formation of the ring system. Previous studies of the RS have focused on optical and UV wavelengths, where the strongest emission lines are H$\alpha$ and Ly$\alpha$, which exhibit boxy profiles \citep{Michael2003,Heng2006,France2010,France2011,Fransson2013,France2015}. The high velocities observed in these lines (reaching up to $\sim 10,000$~\kms), as well as the extent of the emission in images, show that the RS extends from just inside the ER in its plane to well outside it at higher latitudes, though the detailed geometry remains unknown \citep[e.g.,][]{France2015,Larsson2019b}. 

This paper is organized as follows. We describe the observations and data reduction in Section~\ref{sec:obs} and then explain the methods for spectral extraction and construction of 3D emissivity maps in Section~\ref{sec:methods}. The results are presented in Section~\ref{sec:results}, followed by a discussion and conclusions in Sections \ref{sec:discussion} and \ref{sec:conclusions}, respectively. We refer to spectral lines by their vacuum wavelengths throughout the paper.

\section{Observations and data reduction}
\label{sec:obs}

The {\it JWST} NIRSpec IFU observations of SN~1987A were obtained on 2022 July 16 using the G140M/F100LP, G235M/F170LP, and G395M/F290LP gratings/filter combinations.  The three gratings cover the wavelength ranges 0.97--1.88~$\mu$m (G140M), 1.66--3.15~$\mu$m (G235M), and 2.87--5.20~$\mu$m (G235M), with a spectral resolving power ($R=\lambda/\Delta \lambda$) that increases from $R\sim 700$ at the shortest wavelengths to $R \sim1300$ at the longest wavelengths in each grating \citep{Jakobsen2022}.

In order to go as deep in integration as possible over as large a region of sky as possible, we used a small cycling dither pattern with four dithers separated by 0\farcs{25} for the observations. We chose the NRSIRS2RAPID readout mode to take advantage of the IRS2 readout pattern, reducing noise in the data. We kept the number of integrations per exposure at 1 to maximize the signal-to-noise ratio (S/N). The total exposure times were 1751~s for G140M and G235M, and  1225~s for G395M.

In the case of G395M, the exposure time included so-called ``leakcal'' observations, which are used to address the possible problem of light from the sky that leaks though the micro-shutter array (MSA). A detailed description of this issue is provided in Appendix~\ref{app:obs}.  The leakcal observations allowed us to accurately remove the leaked light in G395M. However, we also identified a small region in the G140M and G235M observations that was clearly affected by leakage at wavelengths  $> 1.58\ \mu$m (G140M) and $> 2.65\ \mu$m (G235M).  The affected region is located in the NW part of the ER (indicated in Figure~\ref{fig:fe_image}) and was excluded when extracting spectra. 

We downloaded the observation data from the Mikulski Archive for Space Telescopes (MAST) and ran the data through the Space Telescope Science Institute (STScI) Science Calibration Pipeline.\footnote{https://zenodo.org/record/7504465\#.Y71M47LP1Yh} We used version 1.10.1 of the pipeline.  Full details of the input parameters used for the pipeline are provided in Appendix~\ref{app:obs}. We note that the current version of the pipeline is only able to partly correct for cosmic ray artifacts.

The reference files for the NIRSpec flux calibration have been updated multiple times as the calibration has continued to improve. In order to assess the accuracy of the version used for our analysis,\footnote{Calibration Reference Data System context version 1077} we compared the G395M spectra with the shortest wavelengths of the MRS spectra (Jones et al., in preparation). The two instruments overlap in the wavelength range 4.90--5.20~$\mu$m, which includes a weak \ion{H}{1} $10\rightarrow 6$ line from the ER at 5.129~$\mu$m. We find that the fluxes in this line agree to within $\sim 5\%$, which is comparable to the accuracy of the MRS flux calibration \citep[$5.6\pm 0.7 $\ \%][]{Argyriou2023}, though we note that our comparison is limited by a high noise level in the relevant part of the MRS spectrum.

A comparison of the spectra in the $\sim 0.2~\mu$m wide overlap regions between the NIRSpec gratings give a further indication of the uncertainties in the flux calibration. Spectra were extracted from spatial regions without prominent artifacts to perform this comparison. We find that the agreement is within $\sim 10\%$ and $\sim 5\%$ in the G140M/G235M and G235M/G395M overlap regions, respectively. We stress that the analysis performed in this paper does not rely on accurate flux calibration. In particular, for the 3D emissivity maps, we are only interested in the relative intensities between the emission at different velocities in a given line.

\section{Analysis}
\label{sec:methods}

\subsection{Spectral extraction}

\begin{figure}[t]
\plotone{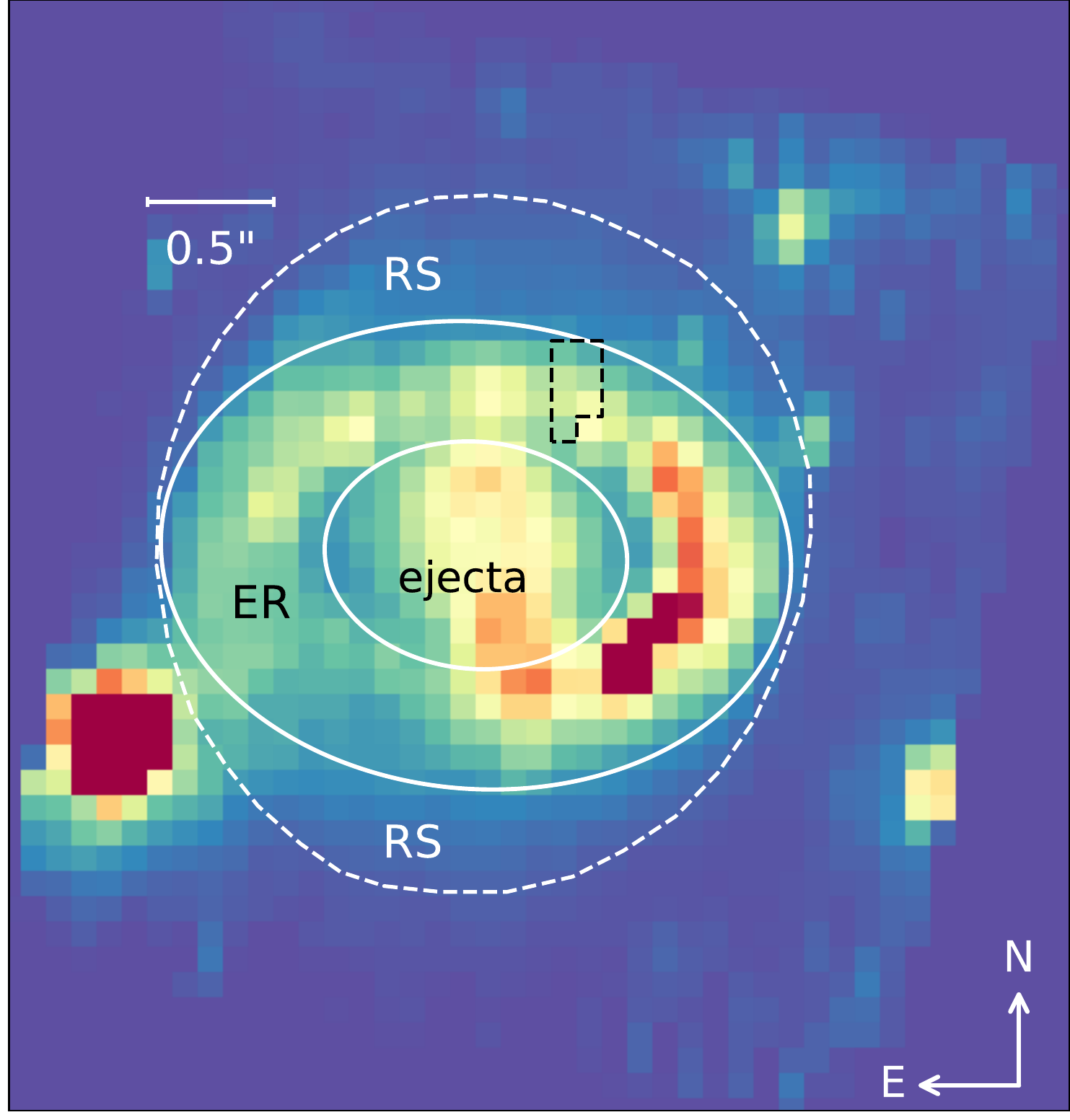}
\caption{NIRSpec image integrated over the wavelength region of [\ion{Fe}{1}]~1.443~$\mu \rm{m} \pm 5000$~\kms. This line originates from the ejecta, while the emission from the ER in this wavelength range is due to narrow [\ion{Fe}{1}]~$1.427~\mu$m and \ion{H}{1}~$1.460~\mu$m lines, as well as continuum. The white solid lines show the regions used to extract spectra for the ejecta (inner ellipse) and ER (elliptical annulus). The white dashed line indicates the approximate area inside which emission from the RS is detected (only a faint continuum component is present at these wavelengths). The dashed black line shows the area affected by light leakage in the long-wavelength ranges of G140M and G235M (see Section~\ref{sec:obs} and Appendix~\ref{app:obs}). The five point sources seen outside the ER are stars. The ER is inclined by 43\dg, with the northern part pointing towards the observer.}
\label{fig:fe_image}
\end{figure}

The NIRSpec cubes have a field of view (FOV) of $3\farcs{3} \times 3\farcs{8}$, which covers the ejecta and ER of SN~1987A. This is illustrated in Figure~\ref{fig:fe_image}, which shows an image of the system produced by integrating the G140M cube over the wavelength region covering the broad [\ion{Fe}{1}]~$1.443~\mu \rm{m}$ line from the ejecta. The figure also shows the regions used for extracting total spectra from the ER and the ejecta. The ER extraction region is an elliptical annulus with semi-major axis 0\farcs{6}--1\farcs{3} and axis ratio 0.75, while the ejecta spectrum was extracted from the elliptical region inside the annulus. The spectra were corrected for the systemic velocity of SN~1987A of 287~\kms\ \citep{Groningsson2008}. 

We extracted background spectra in several small regions outside the ER,  which showed a low and flat background everywhere, with the exception of a mild increase at wavelengths $>5\ \mu$m. However, the size of the background regions is severely limited by the presence of stars, cosmic ray artifacts, extreme-valued pixels, as well as extended emission from SN~1987A itself, which results in poor statistics in the spectra. Because of this and the low background level, we do not subtract a background from the ER and ejecta spectra. 

While the two extraction regions are clearly dominated by the ER and ejecta, respectively, we stress that there is some cross-contamination between the regions. Many of the lines from the ER are so bright that scattered light in the tails of the PSFs contribute significantly in the ejecta region. Conversely, the inner ejecta have now expanded sufficiently to directly overlap with the region of the ER in the south. However, the lines from the ejecta and ER can be distinguished by clear differences in line width, with typical FWHM of $\sim 3500$~\kms\  for the ejecta, compared to $\sim 400$~\kms\ for the ER. Finally, emission from the RS contributes to both extraction regions, as it originates both at the inner edge of the ER and from regions well above and below its plane, which are projected in a large area extending from the center of the ER to up to $\sim 0\farcs{8}$ outside it (see Figure~\ref{fig:fe_image}). Emission from the RS is distinguished by its very broad, boxy profile, extending close to  $\sim 10,000$~\kms\ (see Section~\ref{sec:intro}).

\subsection{Construction of 3D emissivity maps}

For the brightest lines, we can take advantage of the full spatial sampling of NIRSpec and study the spectrum in each spaxel. This is especially interesting for lines originating in the freely-expanding ejecta, for which Doppler shifts and distance from the center of the explosion can be used to create 3D emissivity maps. The assumption of free expansion is expected to hold both for the dense ejecta located inside the ER and the high-velocity ejecta interacting with the RS. The line emission from the RS arises as the ejecta are excited by collisions in the shock region, which does not cause any significant deceleration (though ions are deflected by the magnetic fields, which affects the line profiles of the ionic species; \citealt{France2011}). 

At the time of the observation, 12,927~days after the explosion, one NIRSpec spaxel of 0\farcs{1} (0.024~pc) corresponds to 664~\kms\ in the freely-expanding ejecta, assuming a distance to the LMC of 49.6~kpc \citep{Pietrzynski2019}. This implies that the current semi-major axis of the ER of $0\farcs{82}$ (measured from the hotspots in a recent {\it HST}  image; Larsson et al., in preparation) corresponds to $\sim 5400$~\kms\ for the freely-expanding ejecta, which provides a useful reference point for the velocities. 

We created 3D maps of the bright [\ion{Fe}{1}]~1.443~$\mu$m and \ion{He}{1}~1.083~$\mu$m lines to study the inner metal-rich ejecta and the RS, respectively. The [\ion{Fe}{1}] line is detected at Doppler shifts in the range $[-4000,5000]$~\kms\ and is not expected to be blended with any other strong lines from the ejecta. This is demonstrated by our spectral model in Section~\ref{sec:disc-ejecta}, which shows only minor ($<6$\%) contamination by [\ion{Fe}{1}] lines at 1.437 and 1.462~$\mu$m in the region of the 1.443~$\mu$m line. 

The \ion{He}{1}~1.083~$\mu$m line is detected up to Doppler shifts extending to at least $\pm 8000$~\kms, where emission from other nearby lines becomes significant and makes it difficult to determine the maximal extent of the \ion{He}{1} emission. We focus solely on the RS in the analysis of this line, as the region of  the inner ejecta is expected to have significant contributions from Pa$\gamma$~$1.094~\mu$m and [\ion{Si}{1}]~$1.099~\mu$m in the relevant wavelength region. 

The Pa$\gamma$ line is also expected to have a contribution from the RS, albeit with a much lower flux than the \ion{He}{1} line. To assess the possible contamination by the Pa$\gamma$ RS, we used the Br$\beta~2.626\ \mu$m line, which originates from the same upper level as the Pa$\gamma$ line and can be expected to have the same RS line profile and ratio between the ER and RS emission. We scaled the Br$\beta$ line to the Pa$\gamma$ line using the fact that the peak of the narrow ER component  in the Pa$\gamma$ line is visible above the broad \ion{He}{1} profile.  This comparison shows that contamination from the Pa$\gamma$ RS in the \ion{He}{1} line is expected to become significant at $+7000$~\kms. We therefore limit the analysis of the redshifted \ion{He}{1} emission to $7000$~\kms. 

We assume that the center of the explosion coincides with the systemic velocity of SN~1987A and the geometric center of the ER as determined from {\it HST} observations \citep{Alp2018}. The latter also agrees well with the position of the SN determined from the first {\it HST} observations taken with the Faint Object Camera in 1990 \citep{Jakobsen1991}, when the ejecta were only marginally resolved (Larsson et al., in preparation). The NIRSpec cubes were aligned with a recent {\it HST} image (which had in turn been registered onto {\it Gaia}~DR3) using ``Star 3" to the southeast (SE) of the ER and the brightest star in the northwest (NW) part of the FOV (see Figure~\ref{fig:fe_image}). The FWHM of ``Star 3" at the wavelengths of the \ion{He}{1} and [\ion{Fe}{1}] lines is 0\farcs{18}, which corresponds to $\sim 1200$\ \kms\ for the freely-expanding ejecta. For comparison, the spectral resolving power at the \ion{He}{1} and [\ion{Fe}{1}] lines corresponds to FWHM~$\sim 380$~\kms\ and $\sim 300$~\kms, respectively, implying that the resolution is better along the line of sight in the 3D maps. 

To isolate the ejecta line emission from continuum and contamination by narrow lines from the ER, we performed fits to the spectra of each NIRSpec spaxel in the area of interest. The continuum was determined by fitting a straight line to 500--1000~\kms\ wide intervals on both sides of the emission lines. Wavelength regions contaminated by narrow lines from the ER were fitted with a Gaussian plus a straight line, where the latter is a reasonable approximation of the broad ejecta profile over a limited velocity interval. The fitted velocity intervals cover $\pm$800--1300~\kms\ around the narrow lines ($\sim$5--7 times the resolution at the relevant wavelengths), with larger intervals being used for ER lines that are strong compared to the ejecta. We placed parameter boundaries on the width, central velocity, and normalization of the Gaussian line to ensure that this component did not erroneously fit to substructure in the ejecta profiles. 

In the case of the [\ion{Fe}{1}]~1.443~$\mu$m line, there are two weak lines from the ER in the analyzed wavelength region, [\ion{Fe}{1}]~$1.427~\mu$m and \ion{H}{1}~$1.460~\mu$m. These lines are only detected in and close to the ER, and their removal therefore introduces uncertainties in those regions (i.e., at high velocities in the sky plane in the 3D maps). On the other hand, the narrow lines overlapping with the  \ion{He}{1}~1.083~$\mu$m RS emission are much stronger. The narrow component of this \ion{He}{1} line from the ER is the strongest line in the entire spectrum, and the residuals from subtracting it introduces uncertainties at the lowest Doppler shifts in the entire 3D map of this line. In addition, the narrow component of the Pa$\gamma$~$1.094~\mu$m line discussed above introduces uncertainties at Doppler shifts around $3000$~\kms. 

The cubes produced after subtraction of continuum and narrow lines were inspected for any remaining contamination and other problems. After this inspection, a small number of clear artifacts induced by cosmic rays were masked out in both cubes, and Star 3 was masked out from the \ion{He}{1} cube (it is already outside the analyzed region for [\ion{Fe}{1}]). Finally, both cubes were  linearly interpolated onto a uniform grid with spacing of $100$~\kms\ to aid visualization in 3D.

\section{Results}
\label{sec:results}

\subsection{Spectra of the ejecta and ER}
\begin{figure*}[t]
\begin{center}
\includegraphics[width=20cm,angle=90]{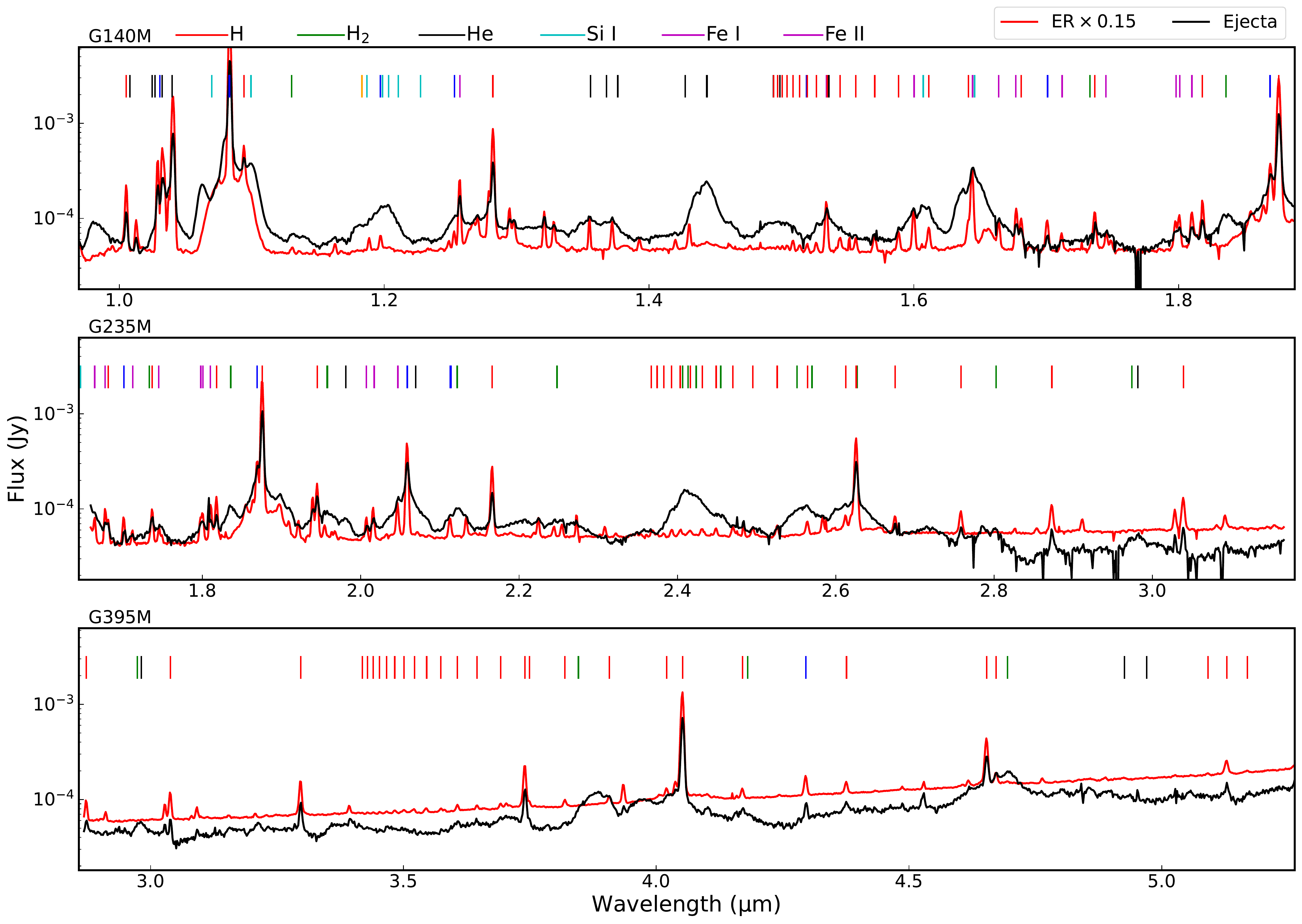}
\caption{Spectra of the ejecta (black) and ER (red) extracted from the regions shown in Figure~\ref{fig:fe_image}. Note that the ejecta region is contaminated by the bright emission from the ER and that some of the ejecta extend into the region of the ER. The strongest lines are identified by the legends, with the color code for different elements and ions given at the top of the figure. All line identifications are also included in Appendix~\ref{app:line_list}. Note the broad, boxy line profiles from the RS, best seen in the \ion{He}{1}~1.083~$\mu$m and Pa$\alpha$~1.875~$\mu$m lines. The ejecta spectrum between $\sim $2.8--3.2~$\mu$m in G235M is highly uncertain due to many artifacts in the form of low-valued spaxels.}
\label{fig:spectra}
\end{center}
\end{figure*}

The full spectra from the regions of the ejecta (black line) and ER (red line) in all three gratings are shown in Figure~\ref{fig:spectra}, with the strongest lines identified. We note the wide dynamic range between the different emission lines and the large difference in flux between the ejecta and ER (the ER spectrum has been multiplied by a factor 0.15 in the figure). The emission from the ER is strongly dominated by the western side (Figure~\ref{fig:fe_image}), as noted in previous MIR and NIR observations \citep{Matsuura2022,Kangas2023}, as well as in optical and soft X-ray images \citep{Larsson2019b, Frank2016}. 

The identifications of atomic lines from the ejecta are based on previous identifications in \cite{Kjaer2010}, as well as on spectral synthesis models discussed in Section \ref{sec:disc-ejecta}. Lines from H$_2$ in the NIR are discussed in \cite{Fransson2016}, and identifications are based on the models in \cite{Draine1996}, discussed in Section \ref{sec:disc-ejecta}. The lines from the ER are narrow and the identifications can be made based on accurate wavelengths in most cases. We have also used shock calculations based on an updated version of the code in \cite{Chevalier1994}. In some cases, especially for weaker, broad ejecta lines, the identifications may be uncertain and should be seen as tentative. Details of all the line identifications are provided in Appendix~\ref{app:line_list}. 

The ejecta lines are discussed in detail in Section \ref{sec:disc-ejecta-results}. Here we just mention some interesting points with respect to the lines from the ER. The spectrum is dominated by a multitude of \ion{H}{1} lines from the Paschen ($n=3$), Brackett ($n=4$), Pfund ($n=5$), and Humphreys ($n=6$) series, and even two high level $n=7$ members (red markers in Figure \ref{fig:spectra}). Furthermore, there are many lines from \ion{He}{1} and \ion{Fe}{2}. 

In addition, we find numerous high-ionization coronal lines from   \ion{Mg}{4}, \ion{Mg}{8}, \ion{Al}{8}, \ion{Al}{9}, \ion{Si}{7}, \ion{Si}{9}, \ion{Si}{10}, \ion{S}{11}, \ion{Ar}{6}, \ion{K}{3}, \ion{Ca}{4}, \ion{Ca}{5},  \ion{Ca}{8}, \ion{Ni}{3}, and \ion{Fe}{13}. Wavelengths and transitions are provided in Table~\ref{tab:coronal_lines} and shown in Figure \ref{fig:coronal_spectra} of Appendix~\ref{app:line_list}. 

High-ionization coronal lines from the ER is no surprise, since they have been observed at different epochs in the optical range \citep{Groningsson2006,Groningsson2008a,Fransson2015}. This is, however, the latest epoch when these high-ionization lines have  been observed, and they can serve as an important diagnostic of the physical conditions in the shocked ER. The optical coronal lines were discussed \cite{Groningsson2006}, using shock models for the ER collision. In their Figure~8, the emissivity of different lines, including the [\ion{Fe}{13}]~$1.075~\mu$m line, were calculated for different shock velocities. This shows that a non-negligible emissivity from ions like [\ion{Fe}{13}] and [\ion{S}{11}] requires a post-shock temperature $\ga 2 \times 10^6$~K, corresponding to a shock velocity $\ga 350$~\kms.

\subsection{3D emissivity of the [\ion{Fe}{1}]~1.443~$\mu$m line from the ejecta}
\begin{figure*}[t]
\plotone{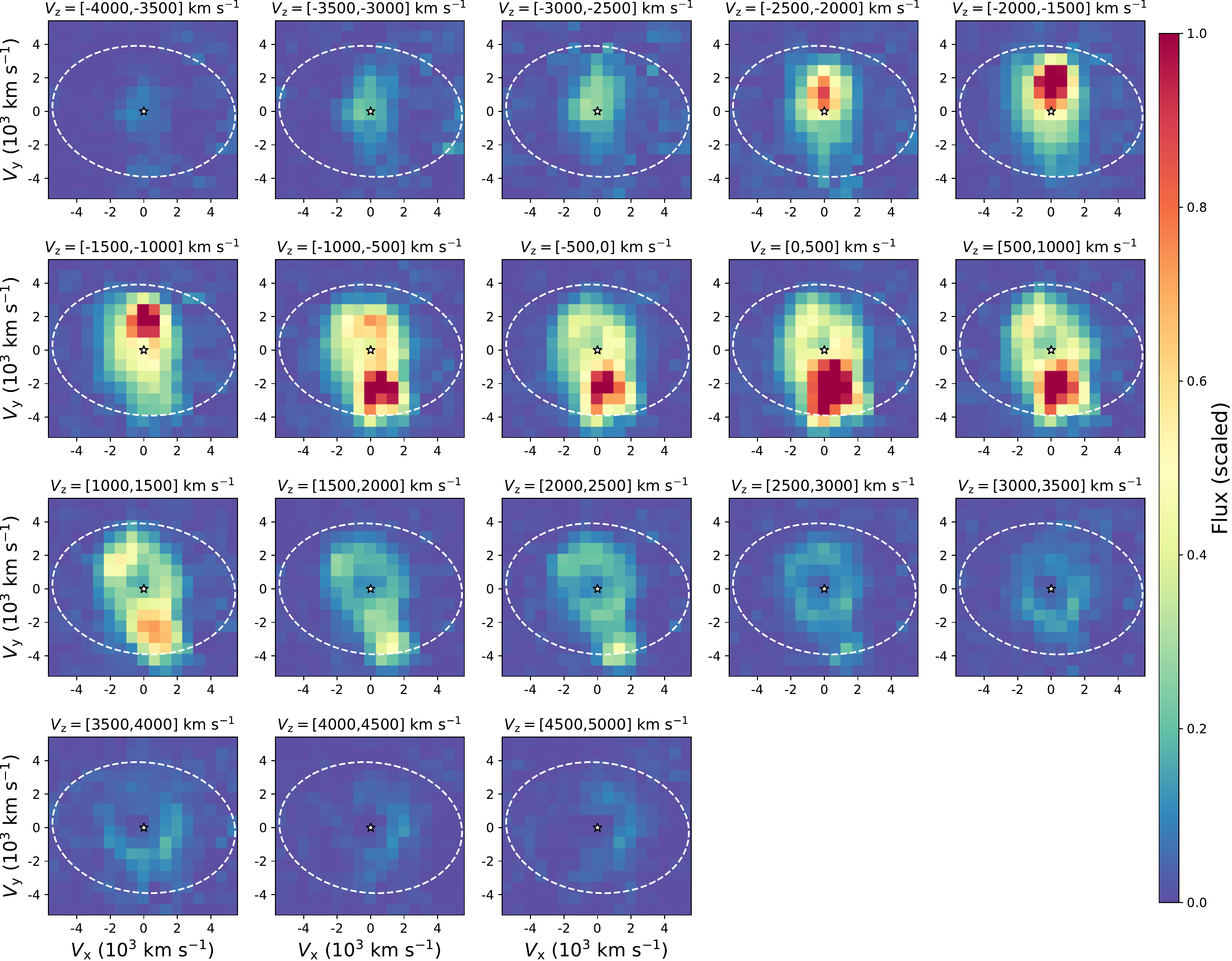}
\caption{Images of the [\ion{Fe}{1}]~1.443~$\mu$m emission from the ejecta as a function of Doppler shift, ranging from $V\rm_z = -4000$~\kms\ (top left) to $V\rm_z =5000$~\kms\ (bottom right). Each image was integrated over an interval of 500~\kms. The velocities of freely-expanding ejecta in the plane of the sky are shown on the x and y axes, where the assumed center of explosion marks 0 velocity (white star symbol). The dashed white line shows the position of the ER.} 
\label{fig:fe_slices}
\end{figure*}
\begin{figure*}[t]
\plotone{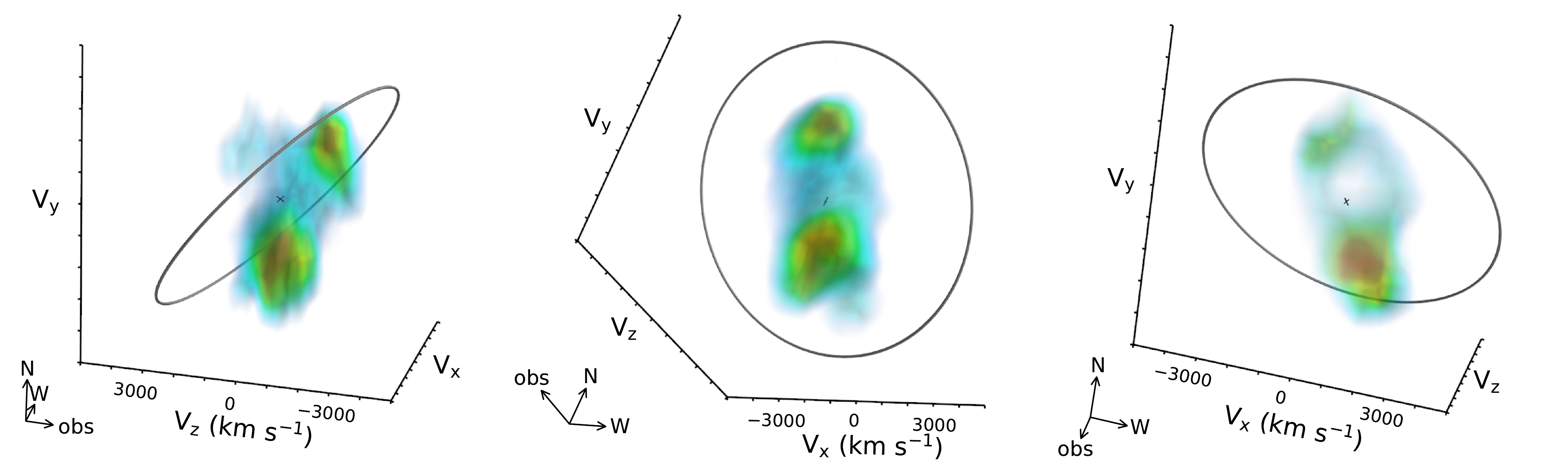}
\caption{Volume rendering of the [\ion{Fe}{1}]~1.443~$\mu$m emission from the ejecta. The three panels show the system from different viewing angles, as indicated by the arrows in the lower left corners. In the left and middle panels, the faintest emission (plotted in blue) corresponds to 15\% of the peak value (red). The right panel only shows ejecta with $V\rm_z >1000$~\kms\ in order to highlight the faint ring of ejecta. In this case, the faintest emission plotted corresponds to 10\% of the peak value. The gray circle shows the inner edge of the ER. An animated version of this figure is available. The video shows one rotation, starting from the viewing angle in the left panel.}
\label{fig:fe_3d}
\end{figure*}

The [\ion{Fe}{1}]~1.443~$\mu$m emission from the ejecta shows an overall elongated morphology along the NE -- southwest (SW) direction ($\sim 15$\dg\ from the north, see Figure~\ref{fig:fe_image}), similar to all other atomic lines previously observed from the ejecta \citep{Wang2002,Kjaer2010,Larsson2013,Larsson2016}. Figure~\ref{fig:fe_slices} shows how this emission is distributed as a function of Doppler shift, after the subtraction of continuum and narrow lines from the ER (Section~\ref{sec:methods}). The spatial scale in the images has been translated to a velocity scale for the freely-expanding ejecta, where $V\rm_x$ denotes velocities in the east-west direction and $V\rm_y$ denotes velocities along the south-north direction.  We use $V\rm_z$ to refer to velocities along the line of sight, where negative velocities correspond to blueshifts. The individual images in  Figure~\ref{fig:fe_slices} were produced by integrating the emission over $500$~\kms\ intervals in $V\rm_z$ between the detection limits of $V\rm_z= [-4000, 5000]$~\kms.

The brightest emission seen in Figure~\ref{fig:fe_slices} is clearly concentrated to a blueshifted clump in the north and a redshifted clump in the south. The 3D space velocities of the peaks of the two clumps are similar, $\sim 2300$~\kms\ in the north and  $\sim 2200$~\kms\ in the south, but their Doppler shifts show that they are not located along the same axis. The peak in the north has a blueshift of $V\rm_z \sim -1500$~\kms, which places it close to the plane of the ER, while the peak in the south has a small redshift of $V\rm_z \sim 100$~\kms. This ``broken dipole" structure is similar to previous SINFONI observations of the [\ion{Fe}{2}]+[\ion{Si}{1}]~1.65~$\mu$m line blend \citep{Kjaer2010,Larsson2016}, but revealed in greater detail in these observations of the  [\ion{Fe}{1}]~1.443~$\mu$m line. 

These NIRSpec observations also reveal features in the ejecta that have not been seen in previous observations. First, it is clear that the inner ejecta are now directly overlapping with the southern part of the ER in the images, as seen from the bright region located at $(V_{\rm x}\sim 1500 ,V_{\rm y} \sim -4000)$~\kms\ at Doppler shifts in the range $V\rm_z = [1500,2500]$~\kms\ (Figure~\ref{fig:fe_slices}). This emission region is above the plane of the ER in 3D, as discussed in Section~\ref{sec:disc-ejecta}. In addition, the images reveal a ring structure in the ejecta that starts to appear at $V\rm_z = -1000$~\kms, but is most apparent on the redshifted side, where faint emission from the ejecta ring can be traced to $V\rm_z = 5000$~\kms. The radius of the ejecta ring is approximately 2.5 spaxels, i.e., $\sim 1700$~\kms, centered near 0 velocity in the sky plane on the redshifted side. The center moves slightly to the north by about one spaxel ($660$~\kms) on the blueshifted side, though this is at least partly influenced by overlapping emission from the bright clump in the south. 

Figure~\ref{fig:fe_3d} shows volume renderings of the [\ion{Fe}{1}]~1.443~$\mu$m emission from different viewing angles, with the inner edge of the ER plotted for reference. This clearly reveals the two main clumps discussed above, where the northern one is approximately in the plane of the ER, while the larger southern clump is below its plane. The figure also shows weak emission between the clumps, as well as on the redshifted side in the north. The ring of ejecta can be seen in the third panel, which only includes ejecta with $V\rm_z >1000$~\kms, while this structure is hidden by the two clumps in the other panels.

\subsection{3D emissivity of the \ion{He}{1}~1.083~$\mu$m line from the reverse shock}

\begin{figure*}[t]
\plotone{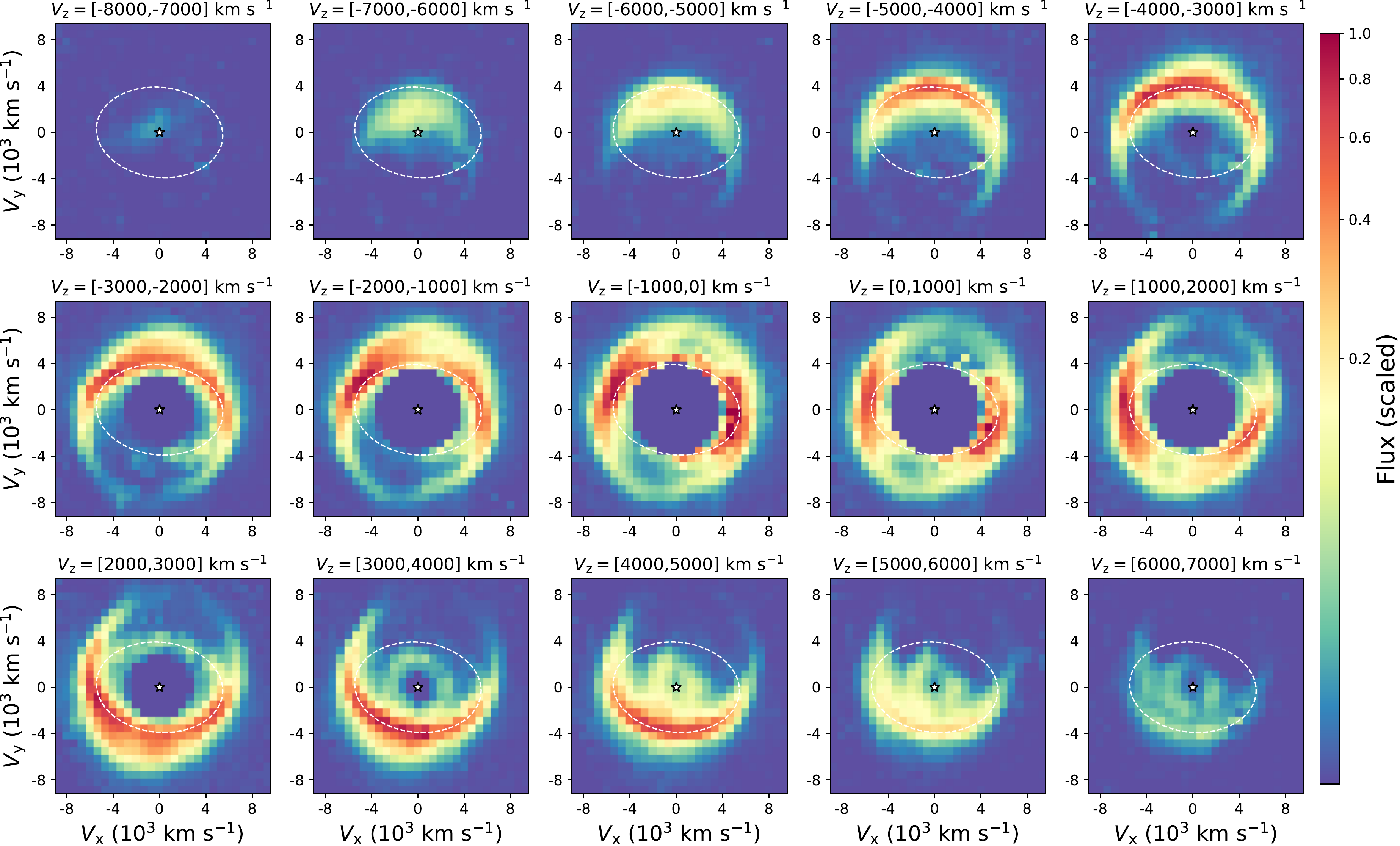}
\caption{Images of the \ion{He}{1}~1.083~$\mu$m emission from the RS as a function of Doppler shift. The \ion{He}{1} emission from the inner ejecta (velocities $<4000$~\kms) has been removed. The last four panels in the bottom row are contaminated by blended emission from Pa$\gamma$~$1.094~\mu$m and [\ion{Si}{1}]~$1.099~\mu$m in the ejecta region. The images have been scaled by an asinh function to show the faintest emission more clearly (see color bar). The dashed white line shows the position of the ER and the white star symbol shows the assumed center of explosion. }
\label{fig:he_slices}
\end{figure*}

Figure~\ref{fig:he_slices} shows images of the \ion{He}{1} emission from the RS as a function of Doppler shift. The images were produced from the cubes after subtraction of continuum and narrow lines, integrating the emission over 1000~\kms\  intervals in $V\rm_z$ between $[-8000, 7000]$~\kms.  The spatial scale was translated to a velocity scale as for the [\ion{Fe}{1}] images in Figure~\ref{fig:fe_3d}. We note that there is some residual contamination from narrow lines from the ER at  $V\rm_z \sim$ 0 and $\sim 3000$~\kms, and that there is contamination by blended emission from Pa$\gamma$~$1.094~\mu$m and [\ion{Si}{1}]~$1.099~\mu$m from the inner ejecta at $V\rm_z>3000$~\kms\ (see Section~\ref{sec:methods}). 

The RS emission is strongest near the ER. As a result, the images in Figure~\ref{fig:fe_3d} show the strongest blueshifted emission in the north and the strongest redshifted emission in the south. Freely-expanding ejecta with velocities $\sim 5400$~\kms\ will have reached the radius of the ER at the time of the observations, which implies that the  \ion{He}{1} emission at higher velocities is expected to originate at high latitudes above and below the plane of the ER. This is also clear from Figure~\ref{fig:he_slices}, which shows highly blueshifted emission projected near the center of the remnant (top left panel). The overall spatial extent of the RS emission has a slight elongation along the NW -- SE direction (by $\sim 15$\dg\ from the north),  which points in the direction of the ORs \citep[e.g.,][]{Crotts2000}, as illustrated below.

\begin{figure*}[t]
\plotone{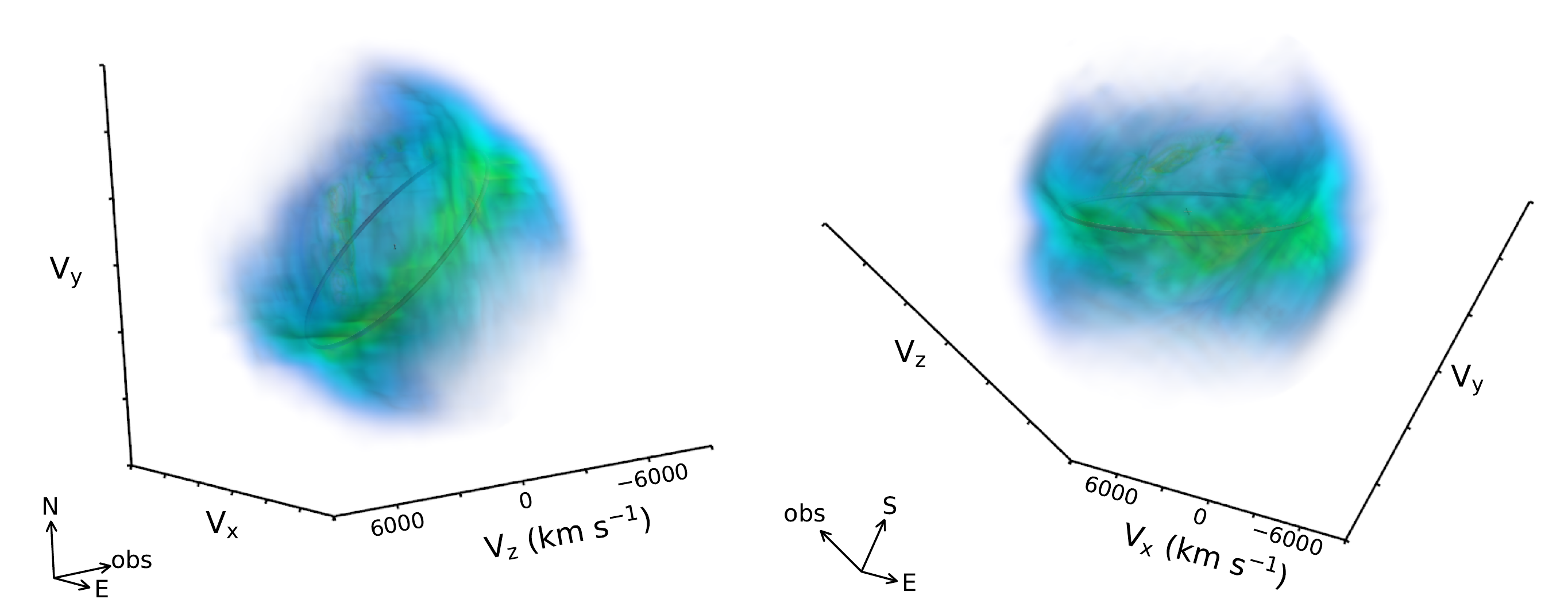}
\caption{Volume rendering of the \ion{He}{1}~1.083~$\mu$m emission from the RS. The two panels show two different viewing angles, as indicated by the arrows in the lower left corners. The emission from the inner ejecta (velocities $<4000$~\kms) has been removed to aid the visualization of the RS. The gray circle shows the position of the ER.}
\label{fig:he_3d}
\end{figure*}

\begin{figure*}[t]
\plotone{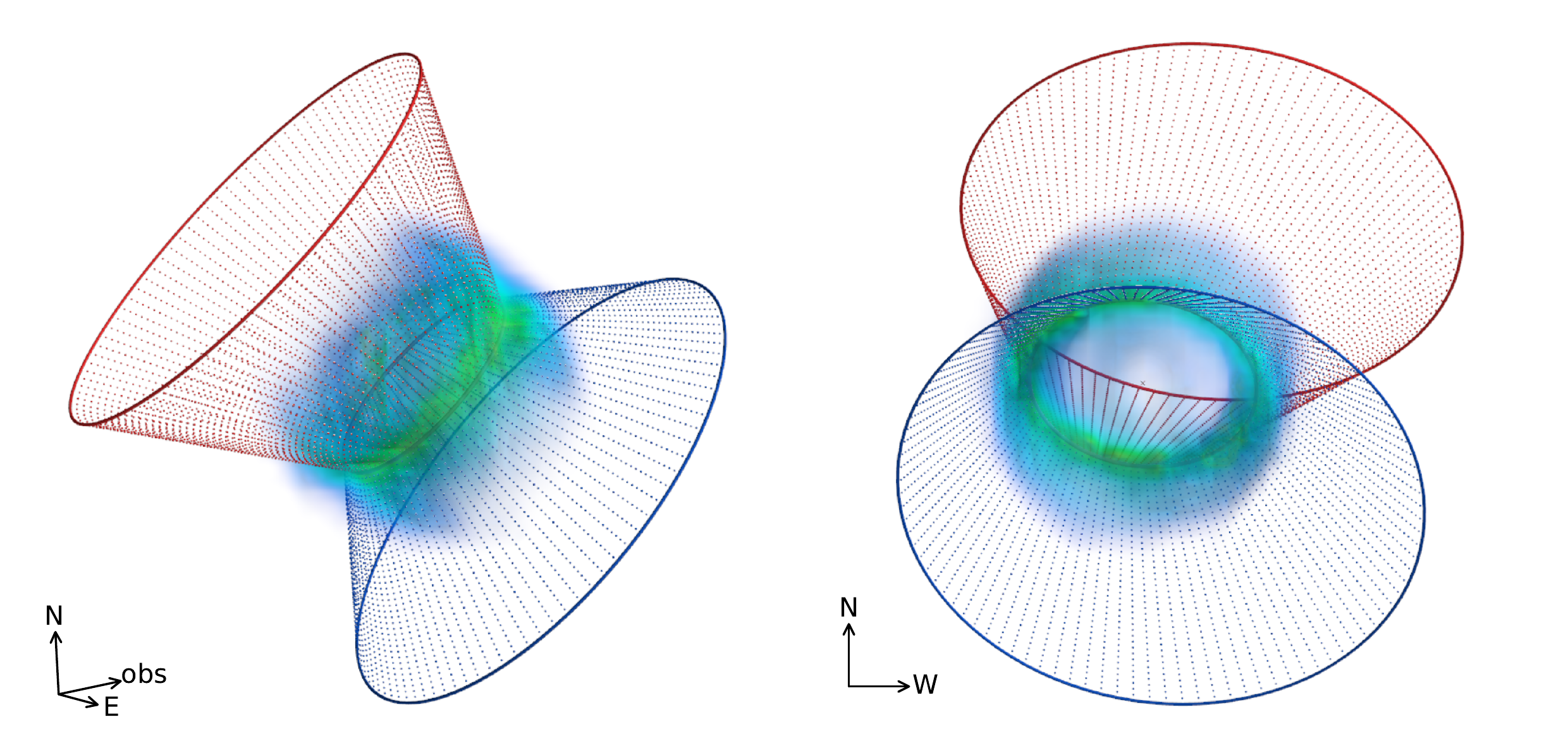}
\caption{Volume rendering of the \ion{He}{1}~1.083~$\mu$m emission from the RS as in Figure~\ref{fig:he_3d}, shown together with the ORs (red and blue circles). The left panel has the same viewing angle as the left panel of Figure~\ref{fig:he_3d}, while the right panel shows the projection on the plane of the sky. The ER is connected to the ORs by dotted lines to aid the visualization.}
\label{fig:he_3d_3ring}
\end{figure*}

A full 3D rendering of the \ion{He}{1} emission is shown in Figure~\ref{fig:he_3d}. This reveals that the emission originates from a surface that extends from the inner edge of the ER to higher velocities on both sides of it with a half-opening angle $\lesssim 45$\dg. The surface then forms a bubble-like structure at higher latitudes above and below the ER. Figure~\ref{fig:he_3d_3ring} shows the RS emission with respect to the ORs. The dimensions and locations of the ORs are the same as described in \cite{Larsson2019b} and verified to agree with the positions of the ORs in a recent {\it HST} image. The ER is connected to the ORs by straight lines in Figure~\ref{fig:he_3d_3ring}, which is the simplest way to connect them. We note, however, that the true geometry of possible walls connecting the ER with the ORs is unknown and likely more complex.  The figure shows that the RS emission is extended in the direction of the ORs (right panel), while the bubble structure at high latitudes is clearly smaller than the simple straight walls connecting the ER with the ORs (discussed further in Section~\ref{sec:disc-he}). 

\begin{figure}[t]
\plotone{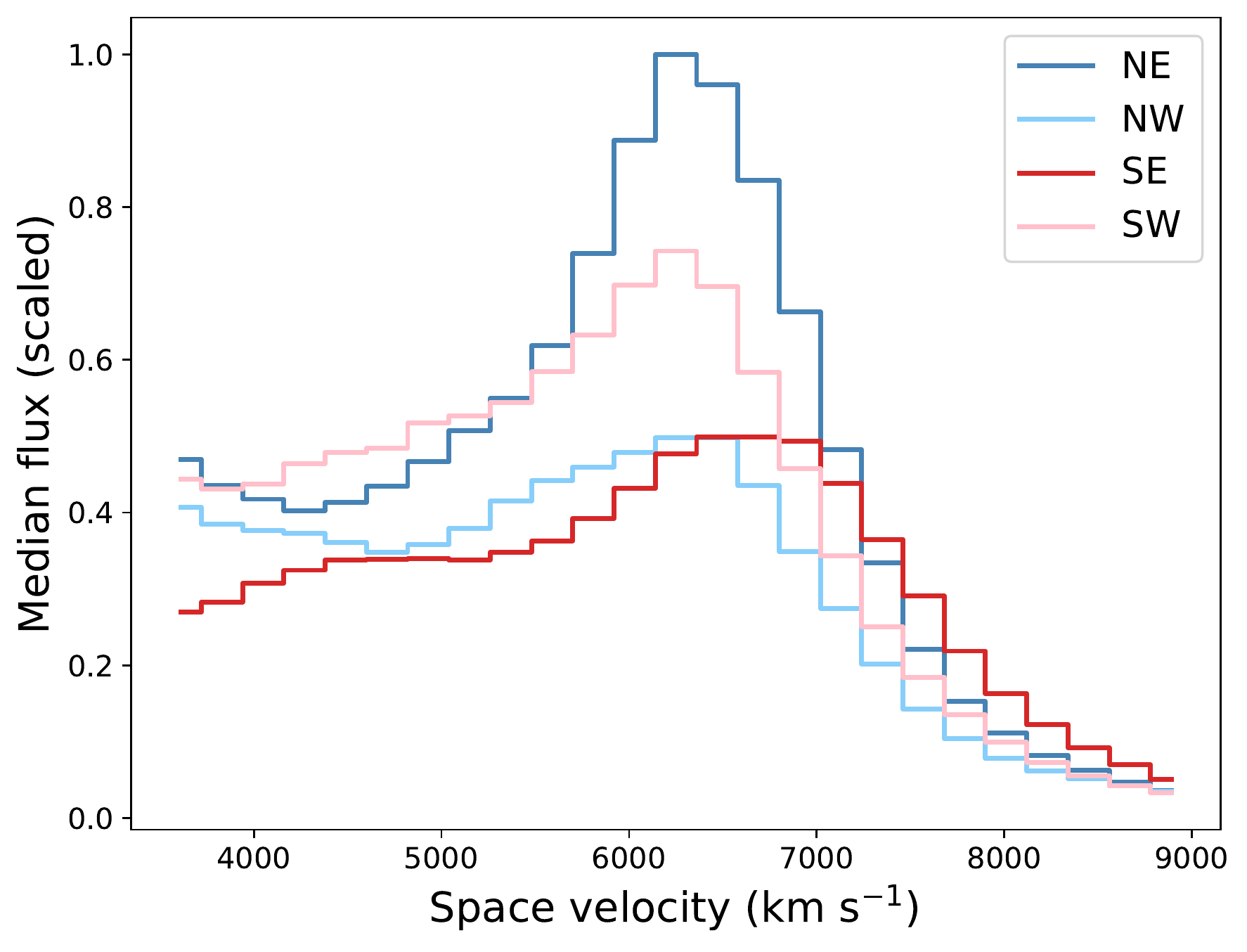}
\caption{Median flux of the \ion{He}{1}~1.083~$\mu$m emission from the RS as a function of the 3D space velocity. The distributions are shown separately for quadrants in the plane of the sky, highlighting the stronger emission in the NE-SW direction and the higher velocities in the SE. }
\label{fig:he_veldist}
\end{figure}

\begin{figure*}[t]
\plotone{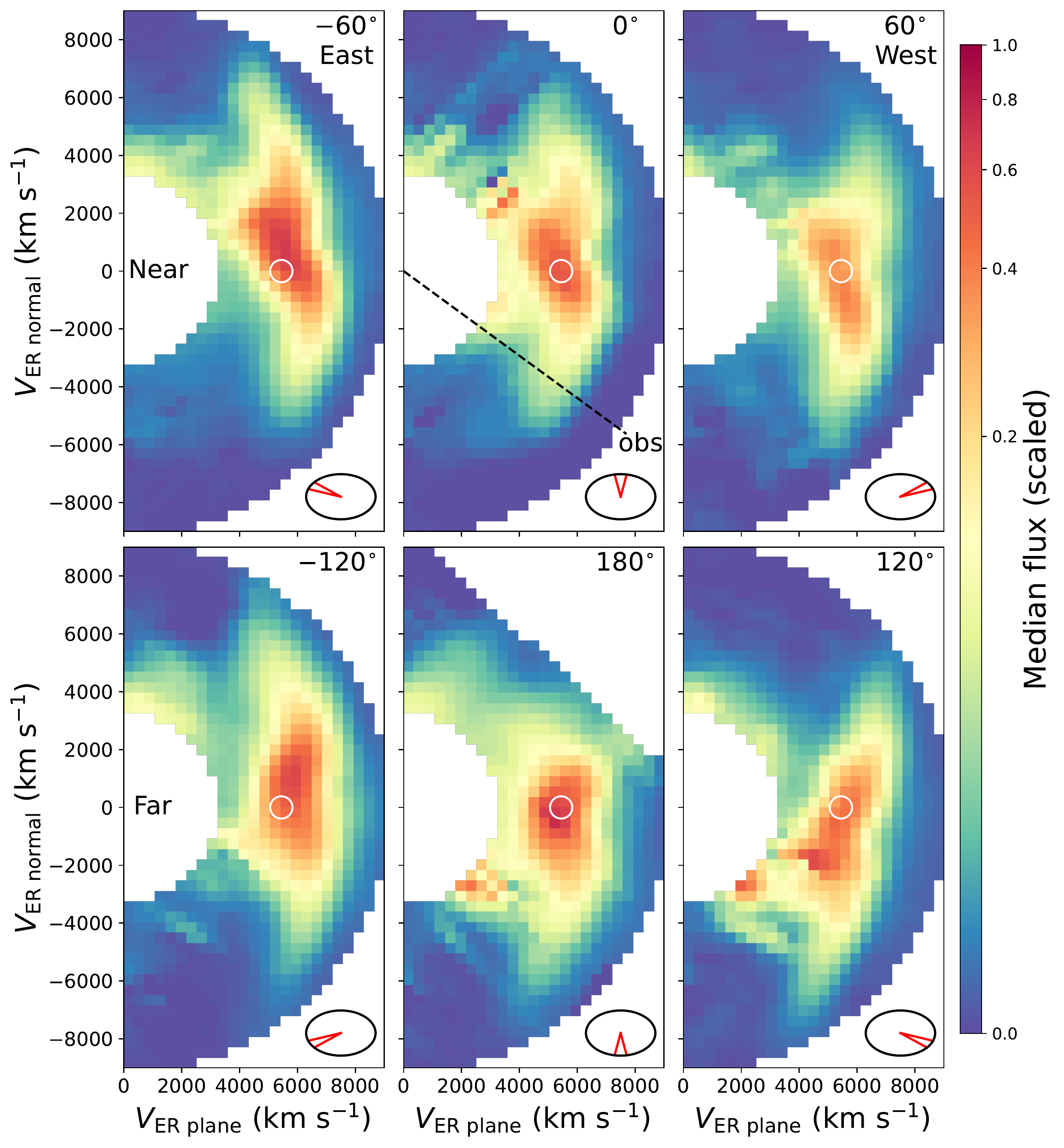}
\caption{Median flux of the \ion{He}{1}~1.083~$\mu$m emission from the RS at different positions along the ER. This quantifies the geometry of the emission region and shows the variations between different positions. The fluxes are shown as a function of the velocity in the plane of the ER (x-axis) and velocity perpendicular to this plane (y-axis), implying that the ER is viewed edge-on. The y-axis is defined such that positive velocities are directed north and away from the observer, while negative ones are directed south and towards the observer. Each ``image" was produced from a $20$\dg\ interval along the ER, centered at the position angles given in the legends, where $0$\dg\ is directly towards the observer. The sketches in the lower right corners show the positions of the segments in the observer frame. Positive (negative) angles are in the western (eastern) parts of the ER. The ER segments included in the top (bottom) rows are on the near (far) side of the ER with respect to the observer. The white circle shows the radius of the ER determined from {\it HST} data. The region covering $V_{\rm ER\ plane} \lesssim 4000$~\kms\ and $V_{\rm ER\ normal}\lesssim \pm 6000$~\kms\ is contaminated by emission from the inner ejecta, dominated by the Pa$\gamma$~$1.094~\mu$m and [\ion{Si}{1}]~$1.099~\mu$m lines. There are also residual artifacts in the ``images" at 0 and 180\dg.}
\label{fig:he_erplane}
\end{figure*}

A more detailed analysis of the RS emission reveals clear evidence of asymmetries, as illustrated by Figures~\ref{fig:he_veldist} and \ref{fig:he_erplane}. Figure~\ref{fig:he_veldist} shows the median flux of the RS as a function of the 3D space velocity, plotted separately for the NE, NW, SE and SW quadrants. This shows that the emission is strongest in the NE-SW direction, i.e., the same direction as the elongation of the inner ejecta. It is also notable that more emission originates at high velocities in the SE, with the peak at $\sim 6700$~\kms\ compared to $\sim 6300$~\kms\ for the other quadrants. The SE region also shows the brightest emission at velocities $\gtrsim 7200$~\kms. 

Figure~\ref{fig:he_erplane} shows the median flux of the RS along two dimensions -- the velocity in the plane of the ER (inclined by 43\dg, \citealt{Tziamtzis2011}) and the velocity perpendicular to this plane. The median was calculated in $220 \times 220$~\kms\ wide intervals in these two velocities, which results in an  ``image" where the 2-dimensional velocity bins define the ``pixels". We performed this calculation for 20\dg-wide segments along the ER, which offers a good balance between obtaining enough statistics in each bin and not blurring the structures too much by combining information from different spatial regions. We show six out of the resulting 18 segments in Figure~\ref{fig:he_erplane}, separated by 60\dg\ to sample locations all along the ER.  This reveals significant spatial variations and deviations from axisymmetry about the normal to the ER plane. To quantify the extent of the main part of the RS emission, we consider contours traced by 20\% of the peak flux, which show that the half-opening angles (defined from the center) are in the range  20--45\dg\ for the different positions. However, it is clear that for the faintest part of the RS emission, the half-opening angle approaches $90$\dg\ in some locations (see e.g.,  the emission below the ER plane at position angle 0\dg\ in Figure~\ref{fig:he_erplane}). The lowest velocity of the RS in the plane of the ER is around 4500~\kms, though this also varies with position. 


\section{Discussion}
\label{sec:discussion}

The NIRSpec observations presented above have, for the first time, provided complete spectra of SN~1987A in the 1--5~$\mu$m range, and allowed us to produce the first  3D emissivity maps of the [\ion{Fe}{1}]~1.443~$\mu$m and  \ion{He}{1}~1.083~$\mu$m lines. We discuss the interpretation of the 3D maps in Sections~\ref{sec:disc-fe} and \ref{sec:disc-he}, respectively.  We then present a basic model for the full ejecta spectrum and discuss the excitation mechanism for the H$_2$ lines in Section~\ref{sec:disc-ejecta}.

\subsection{The asymmetric explosion traced by the [\ion{Fe}{1}]~1.443~$\mu$m line}
\label{sec:disc-fe}
 
The 3D map of the [\ion{Fe}{1}]~1.443~$\mu$m line shows that the Fe produced in the explosion has a similar overall distribution as seen in observations of other atomic lines from the ejecta \citep{Kjaer2010,Larsson2013,Larsson2016,Kangas2022}. The morphology resembles a ``broken dipole" extended along the NW-SW direction, with the emission concentrated near the plane of the ER in the north and closer to the plane of the sky in the south. This geometry is in the same direction as inferred from early observations that probe the outermost ejecta, including polarization {\citep{Schwarz1987,Cropper1988,Jeffery1991}, speckle imaging \citep{Meikle1987,Nisenson1987,Nisenson1999},  the orientation of the ``Bochum" event (transient features in the line profiles, \citealt{Hanuschik1990}) as observed in the spectra of the light echoes  \citep{Sinnott2013}, as well as the location of the first hotspots in the ER \citep{Garnavich1997,Lawrence2000}. The large-scale asymmetries in the explosion hence extend from the inner metal core to the outermost hydrogen envelope. 

An important consideration when interpreting the 3D emissivities is that they do not only reflect the ejecta density, but also the energy sources powering the emission and any obscuration by dust. For the [\ion{Fe}{1}]~1.443~$\mu$m line, the dominant energy source is most likely radioactive decay of $^{44}$Ti, as for the  [\ion{Fe}{2}]+[\ion{Si}{1}]~1.65~$\mu$m line blend \citep{Larsson2016}. However, there may also be significant energy input by the X-ray emission from the ER. This mechanism has been dominating the optical emission for the last two decades \citep{Larsson2011}, and is expected to become increasingly important also for the metal core as the ejecta expand \citep{Fransson2013}.

The X-ray input from the ER may explain the faint ring of ejecta seen in the [\ion{Fe}{1}] emission (Figures~\ref{fig:fe_slices} and \ref{fig:fe_3d}). Due to the steep density gradient at the outer boundary of the metal core, the X-rays are expected to be absorbed in a narrow velocity interval, which would create a ring of emission \citep{Fransson2013}. The appearance of the ejecta ring may also be affected by dust \citep{Matsuura2015}, though ALMA observations show that the spatial distribution of the dust is more elongated than the [\ion{Fe}{1}] ejecta ring \citep{Indebetouw2014,Cigan2019}. A third possibility is that the ejecta ring simply reflects the intrinsic density distribution. Interestingly, a ring/torus of CO emission from the ejecta has been seen in ALMA observations of SN~1987A \citep{Abellan2017}. The radii of the [\ion{Fe}{1}] and CO ejecta rings are similar (1700~\kms), but the CO ring is inclined perpendicular to the ER and hence does not coincide with the [\ion{Fe}{1}] emission. 

Rings of ejecta have also been observed in other young SNRs, including Cas~A \citep{DeLaney2010,Milisavljevic2013,Milisavljevic2015} and SNR~0540-69.3 \citep{Sandin2013,Larsson2021}, indicating that they may be a generic feature, reflecting hydrodynamical instabilities in the explosions 
and/or so-called Ni-bubbles.  The latter arise as the energy input from the radioactive decay of $^{56}$Ni creates low-density regions surrounded by denser walls \citep[e.g.,][]{Li1993,Basko1994,Blondin2001, Gabler2021}, and has been suggested to explain the observed structures in Cas~A \citep{Milisavljevic2015}. Recent 3D simulations show that a large number of Ni-bubbles are created \citep{Gabler2021}, so within this interpretation, the [\ion{Fe}{1}] ejecta ring in SN~1987A may represent the dominant bubble, while other bubbles are too small and/or faint to be detected.

The NIRSpec observations also reveal a bright region of [\ion{Fe}{1}] emission that overlaps with the southern part of the ER as projected on the sky. The Doppler shift of this emission is in the range $V_{\rm z}=[1500,2500]$~\kms\ (Figure~\ref{fig:fe_slices}), which places it above the plane of the ER on the side facing the observer (ejecta directly in the ER plane would have $V_{\rm z} \sim 3800$~\kms). The 3D position of the [\ion{Fe}{1}] region coincides with \ion{He}{1} emission (cf. Figure~\ref{fig:he_slices}), which originates from the RS that extends all along the edge of the southern part of the ER at high latitudes. It is thus likely that the [\ion{Fe}{1}] emission in this region is due to interaction with the RS. 

This is further supported by strong [\ion{Fe}{2}] emission observed from a partly overlapping region. In particular, the [\ion{Fe}{2}] 1.257 and 1.644~$\mu$m lines extracted from the region of the ER show prominent broad features at redshifts $\sim1500$--3000~\kms\ (Figure~\ref{fig:spectra}, top panel). The latter line is blended with [\ion{Si}{1}] 1.646~$\mu$m (discussed in Section~\ref{sec:disc-ejecta-model}), but the absence of a similar redshifted feature associated with the \ion{Si}{1}~1.20~$\mu$m line blend makes it likely that the [\ion{Fe}{2}] dominates at this location. Figure~\ref{fig:fe1p64_red} shows the spatial distribution of this redshifted emission for the [\ion{Fe}{2}] 1.644~$\mu$m line, where the 3D information was obtained using the same methods as for the [\ion{Fe}{1}] and  \ion{He}{1} lines (Section~\ref{sec:methods}). This shows that the [\ion{Fe}{2}] emission extends further to the west (from the south) along the ER compared to the [\ion{Fe}{1}] emission seen in Figure~\ref{fig:fe_slices}, though both lines originate on the near side of the ER with respect to the observer. The [\ion{Fe}{2}] emission from the ejecta also shows signs of interaction at blueshifted velocities near the NE part of the ER, though at a much lower level than in the SW region shown in Figure~\ref{fig:fe1p64_red}. We note that the Fe-rich ejecta in these regions may be excited by X-rays from the RS, rather than by direct shock excitation, which is expected to dominate for \ion{He}{1}.

\begin{figure*}[t]
\plotone{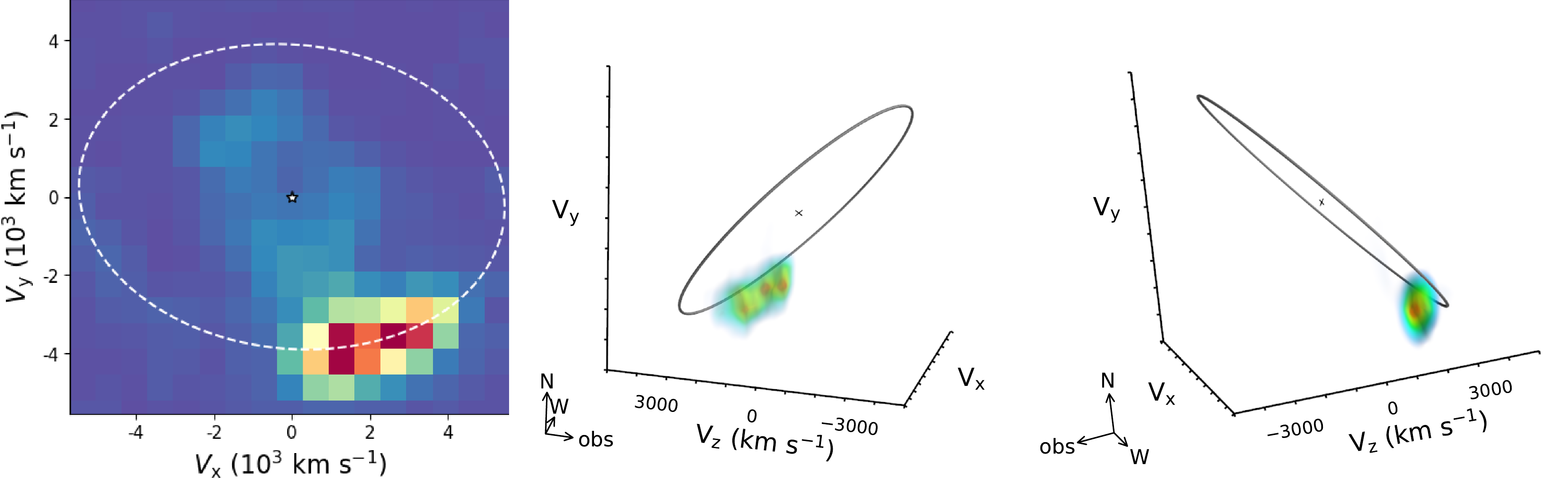}
\caption{Emission from the [\ion{Fe}{2}] 1.644~$\mu$m line from the ejecta at Doppler shifts $V\rm_z >1500$~\kms. The left panel shows an integrated image of this emission, while the middle and right panels show 3D volume renderings from two different viewing angles. The viewing angle of the middle panel is the same as that shown for [\ion{Fe}{1}] in the left panel of Figure~\ref{fig:fe_3d}. The color scale in the image to the left spans the full range of flux values (blue is 0 and red is the maximal flux), while the faintest emission included in the volume rendering (blue)  corresponds to 15\% of the peak value (red).  The white dashed (left) and gray (middle and right) circles show the location of the ER. Note the strong emission located close to the ER in the south, indicating that the dense Fe-rich ejecta are now affected by the shock interaction. An animated version of the 3D rendering is available. The video shows one rotation, starting from the viewing angle in the middle panel.}
\label{fig:fe1p64_red}
\end{figure*}

Evidence of Fe-rich ejecta interacting with the RS has previously been reported based on X-ray observations with {\it XMM-Newton} RGS and EPIC-pn, which show increasing fluxes and centroid energies of Fe~K emission between the years 2010--2019 \citep{Sun2021}. On the other hand, \cite{Maitra2022} find constant Fe abundances in an analysis of {\it XMM-Newton} EPIC-pn and {\it eROSITA} observations, though increases of other elements (like O and Si) are suggested to be due to RS ejecta.
At the same time, {\it Chandra}~HETG observations at soft X-ray energies are consistent with constant abundances up until 2018 \citep{Bray2020,Ravi2021}.  An increasing contribution to the X-ray emission from inner ejecta interacting with the RS has also been predicted by models \citep{Orlando2019,Orlando2020}. The somewhat differing results from the X-ray analyses regarding this suggests that the contribution is still weak, in line with our finding that only a small part of the inner ejecta has reached the RS. 

Further analysis of other emission lines in these NIRSpec observations, including the time-evolution compared to previous SINFONI observations, is expected to make it possible to determine the effects of the energy sources on the observed 3D emissivities. Observational constraints on the 3D distribution of ejecta are important for assessing models for the explosion mechanism, which is a long-standing problem for core-collapse SNe. The leading model for ``ordinary" explosions like SN~1987A is the neutrino-driven mechanism (see \citealt{Janka2017} for a review), while alternatives include explosions powered by jets and/or magnetars \citep[e.g.,][]{Piran2019, Obergaulinger2020}. The latter are likely more relevant for the most energetic SN types, though models involving jets have been proposed also for SN~1987A \citep[e.g,][]{Wang2002,Bear2018}. 

Numerical 3D simulations of neutrino-driven explosions evolved into the remnant stage have shown that the ejecta distribution at late times still retains the imprint of the asymmetries at the time of the explosion \citep{Gabler2021,Orlando2021}. Previous comparisons of SN~1987A  with neutrino-driven explosions have shown that the models can explain the key observables and produce sufficient asymmetries  \citep{Abellan2017,Alp2019,Ono2020,Jerkstrand2020,Utrobin2021,Gabler2021}. However, to date there is no model that also accounts for the complexities of the energy sources and radiative transfer to produce predictions for the optical/NIR emission. It should also be noted that the ejecta asymmetries do not only depend on the explosion mechanism, but are  also affected by the structure of the progenitor star, with most studies of SN~1987A favoring a progenitor that was produced as a result of a binary merger \citep{Menon2017,Menon2019,Ono2020,Orlando2020,Utrobin2021,Nakamura2022}. The CSM geometry as traced by the RS discussed below can give further insight into the evolution of the progenitor of SN~1987A

\subsection{The CSM traced by the \ion{He}{1}~1.083~$\mu$m line from the RS}
\label{sec:disc-he}

The bright \ion{He}{1}~1.083~$\mu$m emission resulting from the interaction between high-velocity ejecta and the RS probes the 3D geometry of the CSM surrounding SN~1987A. The flux depends on the number density of He atoms crossing the shock, so the fact that the emission is strongest in the NE and SW (Figure~\ref{fig:he_veldist}) provides further evidence that the outer ejecta have the same overall distribution as the dense metal core of the ejecta discussed above. 

The spatial distribution of the emission in 3D (Figure~\ref{fig:he_3d}) follows a surface, which confirms that the shock region is narrow. The innermost part of the RS is in the plane of the ER, where the lowest velocity of the emission ($\sim 4500$~\kms, Figure~\ref{fig:he_erplane}) corresponds to $\sim 80\%$ of the ER radius. This position agrees with analytical estimates of ejecta with a steep power-law density profile interacting with a constant-density shell \citep{Chevalier1989}.  Tracing the surface of the RS to higher latitudes shows that the strongest emission is confined to a region with half-opening angle $\lesssim 45$\dg\ around the ER. This is compatible with the thickness of the emission region inferred from {\it HST}/STIS observations of Ly$\alpha$ \citep{Michael2003} and radio observations \citep{Ng2013,Cendes2018}. 

While the overall geometry of the RS is similar all around the ER, there is evidence for asymmetries on a more detailed level. In particular, the SE region displays interaction at higher velocities (Figure~\ref{fig:he_veldist}), which implies that the RS has propagated further in that direction, indicating a lower density of the CSM. This is consistent with the observation that the optical emission from the shocked gas in the ER is faintest in the SE and also peaked at an earlier time ($\sim 7200$ days, compared to $\sim 8300$ in the SW; \citealt{Larsson2019b}). 

The high-latitude RS emission also shows a slight rotation pointing in the direction of the ORs, which may indicate that the CSM near the ER is part of a structure connecting the three rings. The orientation toward the ORs has been observed in previous {\it HST} imaging of the H$\alpha$ emission \citep{France2015,Larsson2019b}, but the NIRSpec observations of the \ion{He}{1} line reveal the 3D geometry of the high-latitude material for the first time. This shows that the RS surface curves inwards at the highest latitudes (Figure~\ref{fig:he_3d} and \ref{fig:he_3d_3ring}), forming a small bubble-like structure, rather than walls of CSM that connect the ER with the ORs in a conical or hour-glass shape, as has previously been discussed \citep[e.g.,][]{Burrows1995,Crotts1995}. For reference, the diameter of the bubble at the greatest distance below the plane of the ER is approximately half that of the simplest model for the walls at the same latitude (Figure~\ref{fig:he_3d_3ring}, left panel, where the walls are just straight lines connecting the ER with the ORs). 

The properties of the CSM between the rings have implications for the formation of the ring system and the pre-SN mass loss in general.  The main properties of the three rings have been explained in a model where they were ejected as a result of a binary merger \citep{Morris2007,Morris2009}, though it is notable that this model does not predict any material located between the rings. An alternative model is that the rings formed as a result of a fast BSG wind interacting with the slower wind from the red supergiant (RSG) phase \citep{Blondin1993,Martin1995,Chevalier1995}. This is predicted to create a bipolar bubble-like geometry, but this model is disfavored by the fact that the ORs  are distinct structures and not the limb-brightened edges of the bubble \citep{Burrows1995}. In addition, this model does not account for the strong mixing in the ejecta needed to explain the CNO abundances, for which a binary merger is a more natural explanation \citep{Fransson1989,Lundqvist1996,Maran2000,Menon2017}. 

Nevertheless, an interesting aspect of the interacting wind scenario is that ionization by the BSG of the swept-up RSG wind may create an \ion{H}{2} region inside the ring system, which is proposed to form a bubble-like structure at high latitudes \citep{Chevalier1995}. The presence of an \ion{H}{2} region in the plane of the ER is supported by early radio and X-ray observations, but the model predictions for the density and location of the interior bubble at high latitudes are uncertain. In the alternative scenario where the rings were created in a merger, the current observations of the RS may instead be probing the mass loss of the progenitor after the merger. The post-merger star may have had an asymmetric wind with a lower density in the polar direction, as may be expected in the case of rapid rotation, explaining the much weaker emission in this direction (Figures~\ref{fig:he_3d}, \ref{fig:he_erplane}).  

The new details of the CSM geometry revealed by the NIRSpec observations highlight the need for more detailed models for the pre-SN mass loss in SN~1987A. The RS is also detected in several other lines in the NIR, (see e.g., the very broad line profiles of several \ion{H}{1} lines in Figure~\ref{fig:spectra}). Further analysis of these lines, together with analysis of the time evolution of the RS in H$\alpha$, will allow us to determine the density and mass of the high-latitude CSM, which will further constrain the formation scenario for the rings and the nature of the progenitor of SN~1987A.

\subsection{Spectral modeling}
\label{sec:disc-ejecta}

The presence of H$_2$ in the ejecta of SN~1987A was first reported by \cite{Fransson2016}, based on observations by VLT/SINFONI in the K band. This was the first detection of H$_2$ in the ejecta of a SN, which confirmed model predictions for Type II SNe \citep{Culhane1995}, and offered a new probe of the physical conditions in the ejecta. The clearest detections in the SINFONI spectra were the 2.40--2.43~$\mu$m and 2.12~$\mu$m lines. Two possible excitation mechanisms for these lines were discussed in \cite{Fransson2016};  fluorescence by UV emission in the 900--1100~\AA \ range, and non-thermal excitation by fast electrons \citep{Gredel1995}, but no firm conclusions could be drawn. The NIRSpec observations allow us to substantially improve our understanding of the H$_2$ emission, owing to the much wider wavelength coverage, in addition to better spatial resolution and S/N. We use simplified spectral models to investigate the excitation mechanism of the H$_2$ lines below. This is combined with models for the atomic lines and continuum emission to create a full model for the ejecta spectrum.  

\subsubsection{Atomic and molecular model}
\label{sec:disc-ejecta-model}

Although the H$_2$ models by \cite{Culhane1995} were pioneering, they did not make specific predictions for the NIR and MIR lines and to date there have not been any updates. For a comparison with our observations, we therefore have to rely on models made for similar physical conditions and excitation mechanisms, in particular models for photodissociation regions (PDRs). While the UV flux in these models has a different origin from that in the SN ejecta, the details of the spectrum in the 900--1100 \AA \ range are found to be of minor importance \citep{Draine1996}. The temperature profiles assumed in these models are also different from that found in the SN ejecta, where the temperature varies considerably between the different abundance zones \cite[e.g.,][]{Jerkstrand2011}. However, the PDR models will at least qualitatively provide important information about the excitation mechanism. \cite{Draine1996} give detailed results for 26 different PDR models with different density, temperature and UV flux parameters,\footnote{https://www.astro.princeton.edu/\~{}draine/pdr.html} which we test using a $\chi^2$ minimization for the strongest lines (the results are presented in Section~\ref{sec:disc-ejecta-results} below).

We also construct a simple model for the atomic lines from the ejecta. As long as ${}^{44}$Ti dominates the energy input to the ejecta, the spectrum changes very slowly (mainly due to the expansion), and the model by \cite{Jerkstrand2011} can be used. However, as discussed in \cite{Fransson2013} and above, the increasing X-ray input from the shock interaction with the CSM will both ionize and heat the ejecta. The affected regions of ejecta now also include the metal-rich zones, as illustrated  by the bright [\ion{Fe}{1}] clump close to the RS in Figures~\ref{fig:fe_slices} and \ref{fig:fe_3d}. This will change the thermal and ionization structure fundamentally, from being powered from the inside by radioactivity, to being powered from the outside by energetic X-rays. Although a fully-consistent model is outside the scope of this paper, we have simulated the spectrum with the most important ions included. 

Our model for the atomic lines assumes a single zone with a given temperature and ionization, here set to 2000~K. This temperature is higher than found for most zones in the purely ${}^{44}$Ti powered case discussed in \cite{Jerkstrand2011}, but may be more typical for the X-ray powered case (see Section \ref{sec:disc-ejecta-results}). Because the spectra of H, He, and Si are dominated by recombination, the \textit{relative} fluxes of the different lines are, however, not very sensitive to the assumed temperature. 

Hydrogen recombination lines are taken from \cite{Hummer1987} and the \ion{He}{1} lines are calculated using the model atom from \cite{Benjamin1999}. The other important ions are \ion{Si}{1}, \ion{Fe}{1}, and \ion{Fe}{2}, originating mainly from the Si/S- and Fe-rich zones in the core, which are now close to the RS. For \ion{Fe}{1}, we use a model atom including 121 levels, for \ion{Fe}{2} 191 levels, and for \ion{Si}{1} 56 levels.  The Fe atoms are similar to the ones used in \cite{Kozma1998}, while we have updated the \ion{Si}{1} atom by including recombination rates to individual levels from \cite{Nahar2000}. Note, however, that these may be too low by a factor of $\sim 2$ below $10^4$~K \citep{Abdel-Naby2012}. We  stress that also the \ion{Si}{1} collision rates are very uncertain. For each ion, the normalization of the flux is set to give a best fit to the lines. The model is therefore a test of the general conditions and line identifications, and not detailed abundances. 

The local line transfer is treated by the Sobolev approximation, assuming free expansion, while the continuum is assumed to be optically thin. We neglect non-thermal excitation in the simulation, although non-thermal ionization is adding to the X-ray ionization. The line profile of the [\ion{Fe}{1}] 1.443 $\mu$m line is the least contaminated strong line in the spectrum, and is therefore used as a template for the integrated line profiles from the ejecta.   

Because of the many overlapping lines, there is no very clear continuum in the spectrum, although there are regions where it may be seen. However, without a continuum, the model gives a bad fit to the spectrum and predicts line profiles that are too peaked. We assume that the continuum is dominated by synchrotron emission from the blast wave and RS. As shown in Figures~\ref{fig:he_slices} and \ref{fig:he_3d}, the RS at high latitudes is projected onto the ejecta, and will therefore contribute to the observed ejecta spectrum.  The synchrotron spectrum is assumed to be similar to that determined from ALMA observations, where \cite{Cigan2019} find a power law $F_\nu \propto \nu^{\alpha}$ with $\alpha = -0.70 \pm 0.06$ for the integrated emission.

In addition to synchrotron emission, we expect H and He continuum emission from both the ejecta and the shocks. To model this, we have included bound-free and free-free emission from \ion{H}{1}, \ion{He}{1}, and \ion{He}{2}, using data from \cite{Ercolano2006} and \cite{vanHooff2014}, respectively. Two-photon emission from \ion{H}{1} and \ion{He}{1} is added from fits by \cite{Nussbaumer1984} and \cite{Schirmer2016}, respectively. For the bound-free and free-free emission we assume a temperature of $10^4$~K and a helium to hydrogen ratio of He/H$=0.17$ by number \citep{Mattila2010}. Above $\sim 3.2 \ \mu$m, there is a clear rise in the continuum level, and we add a simple power law to describe this.

\subsubsection{Synthetic spectra}
\label{sec:disc-ejecta-results}
Figure~\ref{fig:spec_simul} shows the best-fitting complete model for the ejecta, including H$_2$ lines, atomic lines, and continuum. The different contributions to the spectrum are plotted in the lower panel, with the especially important H$_2$ emission shown in blue. The spectra from the three NIRSpec gratings were combined such that  G235M is used in the overlap region with G140M (from $1.70\ \mu$m) due to its higher sensitivity, and G395M in the overlap region with G235M  (from 2.87~$\mu$m) due to many low-valued spaxels at the longest wavelengths in G235M. We note that there are some uncertainties in the flux calibration (Section~\ref{sec:obs}), but this should have a minor effect as we only aim to reproduce the overall appearance of the spectrum.   
\begin{figure*}[t]
\begin{center}
\includegraphics[width=22cm,angle=90]{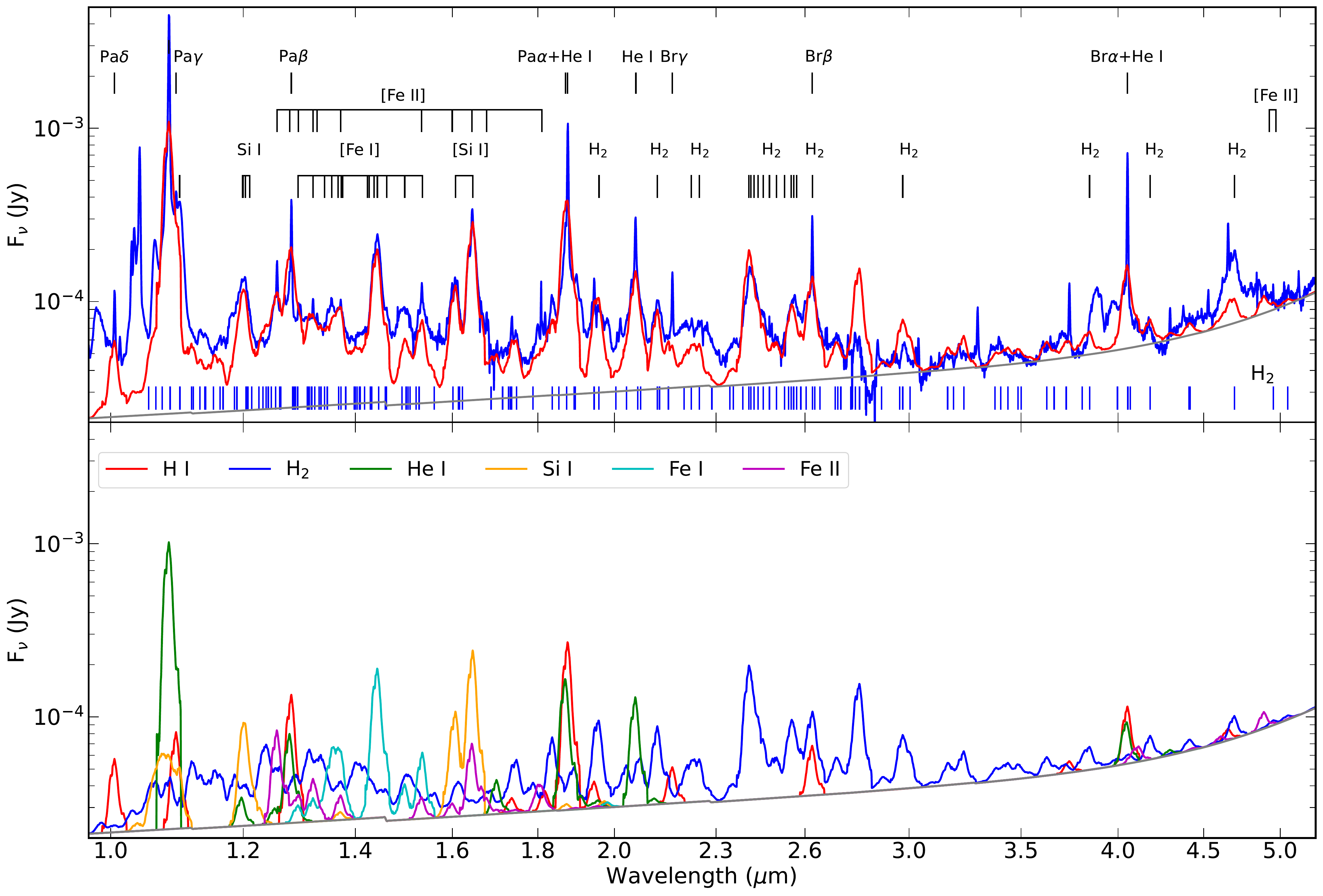}
\caption{Upper panel: Comparison between the NIRSpec ejecta spectrum (blue) and a spectral model (red, see text for details), including \ion{H}{1}, H$_2$, \ion{He}{1}, \ion{Si}{1}, \ion{Fe}{1}, and \ion{Fe}{2}. The gray line shows the continuum, including synchrotron emission, as well as free-free, bound-free and two-photon continua from \ion{H}{1}, \ion{He}{1} and \ion{He}{2}. Lower panel: Contribution of the different ions to the spectrum in the upper panel. Note the large number of H$_2$ lines from the PDR model \texttt{Qm3o} from \cite{Draine1996}. Note also that we have not included any line emission from the RS, which is most clearly seen in the very broad wings of the \ion{He}{1}~1.083~$\mu$m  and Pa$\alpha$ lines. Scattered emission from the ER adds narrow components in some lines, which are especially strong for \ion{H}{1} and \ion{He}{1}. The lines at $\sim 1.035 \ \mu$m may be a blend of \ion{S}{1} and \ion{Fe}{1}, while the feature at $\sim 3.88 \ \mu$m coincides with lines of the same atoms. These lines come from high levels, not included in our model atoms.
}
\label{fig:spec_simul}
\end{center}
\end{figure*}

The model in Figure~\ref{fig:spec_simul} shows that the relative fluxes of the \ion{H}{1} and \ion{He}{1} lines agree well with the observations, as is expected when recombination is dominant. The main source of the ionization of these lines is currently the X-ray input from the interaction with the CSM \citep{Larsson2011,Fransson2013}. 

The total continuum model is shown by the gray line in Figure~\ref{fig:spec_simul}. From the fluxes of the Paschen and Brackett lines, it is clear that the bound-free and free-free continua are too weak to explain the total continuum, although the exact level from the ejecta and ER components is uncertain. Instead, we need a dominant synchrotron component, as discussed above. The slope of this is consistent with the ALMA result \citep{Cigan2019} within the uncertainties. Extrapolating the radio synchrotron continuum from the integrated flux fit in \cite{Cendes2018}, which also fits the ALMA observations in \cite{Cigan2019}, results in a factor of $\sim 7$ higher flux than our adopted continuum. The fraction of the total synchrotron continuum from the RS which falls within the projected area of the ejecta is not known, but should be much less than that coming from the region close to the ER. We therefore believe that this level is very reasonable, and may in fact give a rough estimate of the contribution from the radio emission from high latitudes above the ER. Another caveat is that possible breaks in the synchrotron spectrum between the radio and IR cannot be excluded. 
The rising continuum above $\sim 3.2 \ \mu$m is likely due to hot dust emission.  The origin of this may be either the ejecta-ER collision or the RS.  

Among the PDR models for the H$_2$ lines, it is the  \texttt{Qm3o} model from \cite{Draine1996} that gives the best fit, closely followed by the \texttt{Rh3o}, \texttt{Qw3o} and \texttt{Rw3o} models, with densities $10^4$--$10^6$~cm$^{-3}$ and temperatures 500--1000~K. The fact that these models give the best fit is interesting because, among the models by Draine \& Bertoldi, they have the highest density and UV flux parameter, $\chi = S_{\rm UV}/(4 \pi r^2)$, where $S_{\rm UV}$ is the number of UV photons per second emitted in the 912--1100~\AA \ range, and $r$ is the size of the region. This result illustrates the need for a high UV flux in the ejecta. The main candidates for this are the \ion{He}{1} and \ion{He}{2} two-photon continua from the ejecta, which originate both from radioactive powering and from X-ray input. In addition, the strong continuum and line emission from the ER will be able to penetrate deep into the ejecta and contribute to the ionization. We therefore conclude that UV fluorescence is the main source of excitation for the H$_2$. 

This does not, however, exclude a significant contribution from non-thermal excitation. As shown by \cite{Gredel1995}, for an ionization fraction $\ga 10^{-4}$, the non-thermal excitation and UV-fluorescence models give similar results for the relative line ratios. There are also several sources of fast particles to produce the non-thermal excitation. In particular, thermalization of the high-energy positrons and gamma rays from the ${}^{44}$Ti decay result in fast 10--30~keV secondary electrons in the ejecta \citep{Kozma1992}. In addition, photoelectric absorption in the ejecta of the X-rays from the ER will result in electrons with similar energies. To determine the relative importance of these processes, a detailed model of the H$_2$ excitation for the specific density, temperature, and UV/X-ray/positron source would need to be calculated. 

Regarding the other atomic lines, the ion predominantly responsible for the strong and important [\ion{Fe}{2}]+[\ion{Si}{1}]~1.65~$\mu$m blend  has frequently been discussed. Using the [\ion{Fe}{1}] 1.443\ $\mu$m line as a template, we have compared this line profile to the $\sim 1.65 \ \mu$m feature, assuming either the [\ion{Si}{1}]\ 1.6459\ $\mu$m or the [\ion{Fe}{2}]\ $1.6440\ \mu$m line as the zero velocity reference, and only scaled the absolute flux. We find that the agreement is very good for the [\ion{Si}{1}] line, while there is a systematic blueshift of the profile for the [\ion{Fe}{2}] line. We therefore conclude that [\ion{Si}{1}] dominates the emission, similar to what was concluded at earlier epochs \citep{Jerkstrand2011}.

Among the [\ion{Fe}{2}] lines, the $1.257\ \mu$m line is the least affected by blends.  To reproduce this line, which originates from the same level as the [\ion{Fe}{2}] $1.6440\ \mu$m line, a substantial fraction (approximately one third according to our model, Figure~\ref{fig:spec_simul}) of the $1.60$ and $ 1.65\ \mu$m features are from [\ion{Fe}{2}]. This also gives a good fit to the relative fluxes of the other [\ion{Fe}{2}] lines at 1.257--1.32~$\mu$m and $1.534, 1.600, 1.644, 1.664 ~\mu$m, as is expected since they all arise between the lowest terms, a${}^6$D, ${}^4$F, and ${}^4$D.

The best constraints on the temperature can be obtained from the relative fluxes of the [\ion{Fe}{2}] lines in the NIR and mid-IR. The most sensitive temperature diagnostics are fine-structure lines in the MRS range above $\sim 5.2~\mu$m. However, the \ion{Fe}{2} $4.608, 4.889 ~\mu$m lines from the ground multiplet are in the NIRSpec range, although weak. We find that the observed fluxes of the features at these wavelengths, in combination with the relatively uncontaminated $1.257\ \mu$m line, give a best fit for $\sim 2000$ K, which we assume for the model. We note, however, that this temperature is uncertain, and a more detailed analysis including the MRS data will be discussed in future publications. We also expect the temperature to vary between the inner, X-ray shielded core and the regions close to the RS, like the one in the SW (Figure~\ref{fig:fe1p64_red}). For the line identifications and general conclusions, this assumption is not very important.

The presence of the $\sim 1.20\ \mu$m blend is also interesting. It is well reproduced by \ion{Si}{1} 1.1987--1.2443~$\mu$m lines, which arise from the 4p ${}^3$D level at $\sim 5.96$~eV. This is far above the metastable ${}^1$S and ${}^1$D levels at $\la 1.91$~eV, which give rise to the $1.091$ and $1.607, 1.646 \ \mu$m lines, respectively. While these lines can be populated by collisions, populating the 4p ${}^3$D level by thermal collisions requires a high temperature, $\ga 10^4$ \ K, which would result in even stronger $1.091$ and $1.607, 1.646 \ \mu$m lines and can be excluded. Instead, recombination from \ion{Si}{2} is more likely to be the responsible process. This can occur at  low temperatures and, as shown in the simulation, give a very reasonable ratio between the $\sim 1.20 \ \mu$m lines and the $1.607, 1.646 \ \mu$m lines. The simulation also predicts several other \ion{Si}{1} recombination lines in the 1.06-1.10\ $\mu$m range (Figure \ref{fig:spec_simul}). Unfortunately, this region is swamped by the strong \ion{He}{1} $1.0830 \ \mu$m line from the ejecta and RS, and it is difficult to separate out the \ion{Si}{1} contribution. 

An important caveat with this model is that the atomic data are uncertain for \ion{Fe}{2}, and even more so for \ion{Si}{1}. Also, the highest energy levels of these ions are not included. This is most likely why a few prominent lines in the observed spectrum are not reproduced by the model (e.g., the feature at $\sim 3.9~\mu$m in Figure~\ref{fig:spec_simul}).    
In summary, the first ejecta spectrum covering the full 1--5~$\mu$m wavelength region shows evidence for numerous H$_2$ lines from the H-rich regions, as well as many \ion{Fe}{1}, \ion{Fe}{2}, and \ion{Si}{1} lines from the metal-rich zones, populated by a combination of collisional excitation and recombination. We find that far-UV emission most likely dominates the excitation of the H$_2$. Further self-consistent modeling of the full spectrum will be able to probe the physical conditions in the inner ejecta in more detail.

\section{Summary and conclusions}
\label{sec:conclusions}

We have presented initial results from {\it JWST} NIRSpec IFU observations of SN~1987A, which were obtained as part of GTO program 1232. The observations provide spatially-resolved spectroscopy over the full 1--5~$\mu$m wavelength range for the first time. The IFU makes it possible to disentangle the main emission components of the system: the shocked gas in the ER, the freely expanding inner ejecta, and the high-velocity ejecta interacting with the RS. We have reconstructed the 3D distribution of the [\ion{Fe}{1}]~$1.443\ \mu$m line from the inner ejecta and the \ion{He}{1}~1.083~$\mu$m line from the RS. In addition, we have presented a spectral model for the ejecta, which includes many H$_2$ lines in addition to the atomic lines. 

The [\ion{Fe}{1}]~$1.443\ \mu$m emission shows a highly asymmetric morphology, dominated by two large clumps centered at space velocities of $\sim 2300$~\kms. One clump is located close to the plane of the ER in the north, while the other clump is between the plane of the sky and the ER in the south. This ``broken-dipole" structure resembles previous observations of the [\ion{Fe}{2}]+[\ion{Si}{1}]~1.65~$\mu$m line blend. The NIRSpec observations also reveal a faint ring of ejecta in the [\ion{Fe}{1}] emission, as well as a bright region that directly overlaps with the location of the RS above the plane of the ER in the south, on the side facing the observer. Strong [\ion{Fe}{2}] emission is observed from a partly overlapping region in the SW. This shows that the inner Fe-rich ejecta are now starting to interact with the RS, which will lead to a brightening with time. 

The \ion{He}{1}~1.083~$\mu$m emission has revealed the full 3D geometry of the RS surface for the first time. We find that the RS extends from just inside the ER at $\sim 4500$~\kms\ to higher velocities on both sides of it with a half-opening angle $\lesssim 45$\dg, after which it curves inwards, forming a bubble-like structure at higher latitudes. The RS emission is detected to $\sim 8000$~\kms, with the SE part showing the strongest emission at high velocities, demonstrating clear deviations from axisymmetry. The thickness of the ER as traced by the RS is similar to results from modeling of the radio emission from SN~1987A, but the curvature at higher latitudes has not been seen in previous observations. This calls for more detailed model predictions of the pre-SN mass loss and formation of the ring system.  

Our spectral model of the ejecta aids the identification of the emission lines, many of which are blended, including the numerous H$_2$ lines. We find that the H$_2$ lines are well described by PDR models characterized by a strong UV flux in the 912--1100~\AA \ region. The origin of the UV continuum is likely the two-photon He emission from the ejecta and ER, though more detailed modeling is needed to draw firm conclusions. The metal line ratios from the ejecta are consistent with a combination of collisional excitation and recombination in the low-temperature inner ejecta. 

Further analysis of this data set will provide more detailed information about the physical properties and spatial distribution of both the inner ejecta and the CSM traced by the RS. The observations will also allow us to address a wide range of other topics, including the properties of dust and shocked gas in the ER, as well as possible emission from a compact object. These studies, together with analyses of the  MIRI MRS and Imager data from this GTO program, will thus greatly improve our understanding of this historic event.

\begin{acknowledgments}
This work is based on observations made with the NASA/ESA/CSA James Webb Space Telescope. The data were obtained from the Mikulski Archive for Space Telescopes at the Space Telescope Science Institute, which is operated by the Association of Universities for Research in Astronomy, Inc., under NASA contract NAS 5-03127 for JWST. These observations are associated with program \#1232. The specific observations analyzed can be accessed via \dataset[DOI: 10.17909/175h-7x33]{https://doi.org/10.17909/175h-7x33}.

JL acknowledges support from the Knut \& Alice Wallenberg Foundation. 
JL and CF acknowledge support from the Swedish National Space Agency. 
OCJ acknowledge support from an STFC Webb fellowship. 
MM and NH acknowledge support through a NASA/JWST grant 80NSSC22K0025, and MM and LL acknowledge support from the NSF through grant 2054178.
JH was supported by a VILLUM FONDEN Investigator grant (project number 16599).
ON acknowledges support from STScI Director's Discretionary Fund.
MJB acknowledges support from European Research Council Advanced Grant 694520
SNDUST.
PJK and JJ acknowledges support the Science Foundation Ireland/Irish Research Council Pathway programme under Grant Number 21/PATH-S/9360.
TT acknowledges financial support from the UK Science and Technology Facilities Council, and the UK Space Agency.  MM acknowledges that a portion of her research was carried out at the Jet Propulsion Laboratory, California Institute of Technology, under a contract with the National Aeronautics and Space Administration (80NM0018D0004).

\end{acknowledgments}

\vspace{5mm}
\facilities{JWST (NIRSpec)}

\software{Astropy \citep{Astropy2022},
    Matplotlib \citep{Hunter2007},
	Mayavi \citep{Ramachandran2011}
          }

\newpage
 
\appendix

\section{Details about the observations and data reduction}
\label{app:obs}

Here we provide additional information about the issue of light leakage through the MSA (Section~\ref{app:obs-leak}) and the input parameters used for the calibration pipeline (Section~\ref{app:obs-calpar}). 

\subsection{Light Leakage}
\label{app:obs-leak}

One issue of concern for the NIRSpec IFU is light from the sky that enters through open shutters in the MSA, as well as light that ``leaks'' through closed MSA shutters.  A means of addressing this problem is to use ``leakcal'' observations, which are observations with the IFU aperture closed of the exact same field as when the IFU shutter is open. This makes it possible to subtract the contribution of light that leaks through the MSA and thereby isolate the light that passed through the IFU aperture.  However, the leakcal observations can take significant observing time to obtain, so it is important to determine exactly in which instances they are required.  

To this end, we inspected observations of the regions including and surrounding SN 1987A obtained by the VISTA survey of the Magellanic Clouds system \citep[VMC; ][]{cioni11} at Y ($\sim$ 1.0 $\mu$m wavelength), J ($\sim$ 1.2 $\mu$m), and K$_{s}$ ($\sim$ 2.2 $\mu$m) bands, and also at bands 1 and 2 ($\sim$ 3.6 and $\sim$ 4.5 $\mu$m wavelength, respectively) of the Infrared Array Camera \citep[IRAC; ][]{fazio04} on the {\it Spitzer Space Telescope} \citep[][]{werner04}.  These observations span roughly the $\sim$ 1-5 $\mu$m wavelength range covered by the NIRSpec IFU observations.  The VMC Y and J observations showed hardly any extended emission near SN 1987A where the MSA quadrants could land.  The K$_{s}$ observation did show some mild extended emission northeast (NE) of SN 1987A.  In the {\it Spitzer}-IRAC 3.6 $\mu$m image, the nebulosity NE of SN 1987A becomes more prominent, and in the {\it Spitzer}-IRAC 4.5 $\mu$m image, this nebulosity is quite prominent.  It was decided that the best way to deal with leakage through the MSA was to apply an Aperture PA range special requirement in the Astronomer$'$s Proposal Tool (APT) on the NIRSpec IFU observation of 190 -- 70\dg.  This would avoid the nearby nebulosity in the K$_{s}$ image, so that no leakcal would be required for the G140M or G235M observations.  The nearby nebulosity for G395M was quite extensive, so we decided to obtain leakcals at all dither positions for the G395M observation.

When the observations were obtained, we inspected the G140M, G235M, and G395M data (both science target and leakcal data for G395M).  The G395M leakcal clearly shows emission that did not enter through the IFU aperture, validating our decision to obtain leakcal observations to isolate and remove it.  Without leakcal observations for G140M or G235M, it is not entirely straightforward to determine what signal in these gratings came through the MSA outside of the IFU aperture. However, we can identify locations in this data where such signal is strongest and corresponds to signal in the G395M leakcal data.  We find that the leakage primarily affects a small region overlapping with the NW part of the ER at wavelengths $> 1.58\ \mu$m in G140M and $> 2.65\ \mu$m in G235M. This region was excluded when extracting spectra.

\subsection{Parameters for the calibration pipeline}
\label{app:obs-calpar}

When running the Level 1 pipeline, we used the default pipeline parameters except for a few exceptions.  First, we skipped the \texttt{reference pixel} step.  Next, for the \texttt{jump} step, we set expand\_large\_events to True, maximum\_cores to “half", use\_ellipses to False, expand\_factor to 2.5, after\_jump\_flag\_dn1 to 1000, and after\_jump\_flag\_time1 to 90.  Lastly, for the \texttt{ramp\_fit} step, we set maximum\_cores to “half".  We set the parameters for the \texttt{jump} and \texttt{ramp\_fit} steps as we do in order to correct for artifacts we assume are mostly due to cosmic rays.  At present, this correction is only partial.  While some of these artifacts are adequately removed, others are only partly corrected or not at all. With the output from the Level 1 pipeline, if the value of a pixel in the DQ extension is greater than 0, we set the value of that pixel to 1, as this ensures that the data for this pixel would not be used when building the cube in Level 3 of the pipeline.  In Level 2 of the pipeline, we use the default pipeline parameters.  Lastly, in Level 3 of the pipeline, we use the default pipeline parameters except that we skip the \texttt{outlier\_detection} step, as we found that this step was too aggressive in its identification of outlier spaxels.

\section{List of observed lines from the ER and ejecta}
\label{app:line_list}

In this Appendix we provide a list of the lines identified in the spectra, marked in Figure \ref{fig:spectra}. Lines from neutral and singly ionized elements in the ER are included in Table~\ref{tab:erlines}, while the high ionization coronal lines are listed in Table~\ref{tab:coronal_lines} and marked in Figure \ref{fig:coronal_spectra}. To avoid the velocity smearing when extracting the integrated ER spectrum, we have extracted a small $0.13$ arcsec$^2$ region in the south western part of the ER. Wavelengths and uncertainties for the coronal lines are taken from NIST, \cite{Feuchtgruber1997} and \cite{Casassus2000}. The lines from the inner ejecta are included Table~\ref{tab:ejectalines}. 

The identification of narrow lines from the ER are secure in most cases, but there are considerable uncertainties in the identification of the ejecta lines because of blending between the broad lines. For the ejecta, the identification is mainly based on the modeling discussed in Section \ref{sec:disc-ejecta-results}. There are also some cases where we have not found any clear identification, marked with a question mark. There may also be additional coronal lines from the ER, but they are either very weak or have wavelengths outside the quoted error-bars of the laboratory wavelengths. 
\startlongtable
\begin{deluxetable*}{llll}
\tablecaption{Line identifications for the ER}
\tablehead{
\colhead{Wavelength}& \colhead{Ion}& \colhead{Transition}\\
\colhead{($\mu$m)}&&\\
}
\startdata
0.9549& \ion{H}{1}  & 3 -- 8  &  \\
1.005& \ion{H}{1}  & 3 -- 7 &  \\
1.013&?&\\
1.031& \ion{He}{1} & 3p ${}^3$P -- 6d  ${}^3$D  &   \\
1.083& \ion{He}{1} & 2s ${}^3$S -- 2p ${}^3$P  & \\
1.094& \ion{H}{1}  & 3 -- 6 & \\
1.197& \ion{He}{1} &    3p ${}^3$P -- 5d${}^3$D   \\
1.253& \ion{He}{1} &    3s ${}^3$S -- 4p${}^3$P    \\
1.282& \ion{H}{1} & 3 -- 5 &   \\ 
1.257& [\ion{Fe}{2}]&   a ${}^6$D$_{9/2}$ -- a ${}^4$D$_{7/2}$   \\
1.279& [\ion{Fe}{2}]&    a ${}^6$D$_{3/2}$ -- a ${}^4$D$_{3/2}$  \\
1.295& [\ion{Fe}{2}]&    a ${}^6$D$_{5/2}$ -- a ${}^4$D$_{5/2}$  \\
1.321& [\ion{Fe}{2}]&     a ${}^6$D$_{7/2}$ -- a ${}^4$D$_{7/2}$ \\
1.328& [\ion{Fe}{2}]&    a ${}^6$D$_{3/2}$ -- a ${}^4$D$_{5/2}$  \\
1.372& [\ion{Fe}{2}]&     a ${}^6$D$_{9/2}$ -- a ${}^4$D$_{5/2}$ \\
1.519& \ion{He}{1} &    3s ${}^1$S -- 4d ${}^1$D    \\
1.520 & \ion{H}{1}  &  4 -- 20&  \\
1.526 & \ion{H}{1}  & 4 -- 19  &   \\
1.534& [\ion{Fe}{2}]& a ${}^4$F$_{9/2}$ -- a ${}^4$D$_{5/2}$  & \\
1.544& \ion{H}{1}  &    4 -- 17 \\
1.556& \ion{H}{1}  &    4 -- 16 \\
1.570& \ion{H}{1}  &  4 -- 15   \\
1.588& \ion{H}{1}  &  4 -- 14   \\
1.600& [\ion{Fe}{2}]& a ${}^4$F$_{7/2}$ -- a ${}^4$D$_{3/2}$  & \\ 
1.611& \ion{H}{1}  &  4 -- 13   \\
1.641& \ion{H}{1}  &   4 -- 12  \\
1.644& [\ion{Fe}{2}]&a${}^4$F$_{9/2}$ -- a${}^4$D$_{7/2}$ & \\
1.664& [\ion{Fe}{2}]&a$ {}^4$F$_{5/2}$ -- a ${}^4$D$_{1/2}$ & \\ 
1.677& [\ion{Fe}{2}]&a ${}^4$F$_{7/2}$ -- a ${}^4$D$_{5/2}$ & \\  
1.681& \ion{H}{1}  &   4 -- 11  \\
1.701& \ion{He}{1} &   3p ${}^3$P -- 4d ${}^3$D     \\
1.712& [\ion{Fe}{2}]&  a ${}^4$F$_{5/2}$ -- a ${}^4$D$_{3/2}$ \\  
1.737& \ion{H}{1}  & 4 -- 10 & \\ 
1.745& [\ion{Fe}{2}]&  a ${}^4$F$_{3/2}$ -- a ${}^4$D$_{1/2}$    \\
1.798& [\ion{Fe}{2}]&  a ${}^4$F$_{3/2}$ -- a ${}^4$D$_{3/2}$    \\
1.801& [\ion{Fe}{2}]&  a ${}^4$F$_{5/2}$ -- a ${}^4$D$_{5/2}$  & \\
1.810& [\ion{Fe}{2}]&  a ${}^4$F$_{7/2}$ -- a ${}^4$D$_{7/2}$  & \\
1.818& \ion{H}{1} & 4 -- 9& \\ 
1.869& \ion{He}{1} &     3d ${}^3$D -- 4f${}^3$F  \\
1.875& \ion{H}{1}  & 3 -- 4 & \\ 
1.945& \ion{H}{1}  &   4 -- 8  \\
2.007& [\ion{Fe}{2}]&   x ${}^6$P$_{5/2}$ -- ${}^6$P$_{7/2}$   \\
2.017& [\ion{Fe}{2}]&    e ${}^4$G$_{5/2}$ -- ${}^4$D$_{5/2}$ \\
2.047& [\ion{Fe}{2}]&     e ${}^4$G$_{9/2}$ -- ${}^4$G$_{9/2}$ \\
2.059& \ion{He}{1} & 2s ${}^1$S -- 2p ${}^1$P  & \\
2.113& \ion{He}{1} &      3p ${}^3$P -- 4s ${}^3$S \\
2.114& \ion{He}{1} &     3p ${}^1$P -- 4s ${}^1$S  \\
2.166& \ion{H}{1}  & 4 -- 7 &  \\
2.431& \ion{H}{1}  &   5 -- 20  \\
2.449& \ion{H}{1}  &   5 -- 19  \\
2.470& \ion{H}{1}  &  5 -- 18   \\
2.495& \ion{H}{1}  &   5 -- 17  \\
2.526& \ion{H}{1}  &   5 -- 16 \\
2.564& \ion{H}{1}  &  5 -- 15   \\
2.613& \ion{H}{1}  &   5 -- 14  \\
2.626& \ion{H}{1}  &  4 -- 6&  \\
2.675& \ion{H}{1}  &   5 -- 13  \\
2.758& \ion{H}{1}  & 5 -- 12 & \\ 
2.873& \ion{H}{1}  & 5 -- 11 &\\ 
3.039& \ion{H}{1}  & 5 -- 10 & \\
3.297& \ion{H}{1}  & 5 --9 &  \\ 
3.607& \ion{H}{1}  & 6 -- 20 &  \\
3.741& \ion{H}{1}  & 5 -- 8 & \\ 
3.744& \ion{H}{1}  &   ??  \\
3.749& \ion{H}{1}  &  6 -- 17   \\
3.819& \ion{H}{1}  &   6 -- 16  \\
3.908& \ion{H}{1}  &   6 -- 15  \\
3.935&?&\\
4.021& \ion{H}{1}  &  6 -- 14    \\
4.052& \ion{H}{1}  & 4 -- 5 & \\ 
4.171& \ion{H}{1}  &   6 -- 13  \\
4.296& \ion{He}{1} &    3s ${}^3$S -- 3p ${}^3$P   \\
4.376& \ion{H}{1}  &  6 -- 12    \\
4.654& \ion{H}{1}  & 5 -- 7 &\\
4.673& \ion{H}{1}  &   6 -- 11  \\
5.091& \ion{H}{1}  & 7 -- 20 &  \\
5.129& \ion{H}{1}  & 6 -- 10 &  \\
5.135& \ion{H}{1}  &  ??   \\
5.169& \ion{H}{1}  &  7 -- 19   \\
\enddata
\label{tab:erlines}
\end{deluxetable*}

\startlongtable
\begin{deluxetable*}{llll}
\tablecaption{Coronal lines identified in the ER}
\tablehead{
\colhead{Wavelength}& \colhead{Ion}& \colhead{Transition}&\colhead{Note}\\
\colhead{($\mu$m)}&&\\
}
\startdata
1.0750&\ion{Fe}{13}&${}^3$P$_{0}$ -- ${}^4$D$_{1}$&On the blue wing of \ion{He}{1} 1.083 $\mu$m\\
1.3929 &\ion{S}{11}&${}^3$P$_{1}$ -- ${}^3$P$_{2}$&\\
1.4305 &\ion{Si}{10}&${}^2$P$_{1/2}$ -- ${}^2$P$_{3/2}$&\\
1.9213& \ion{S}{11}&${}^3$P$_{0}$ -- ${}^3$P$_{1}$ \\
2.045& \ion{Al}{9}&${}^2$P$_{1/2}$ --  ${}^2$P$_{3/2}$&\\
2.3211&\ion{Ca}{8}&${}^2$P$_{1/2}$ -- ${}^2$P$_{3/2}$&\\
2.481& \ion{Si}{7}&${}^3$P$_{2}$ --  ${}^3$P$_{1}$& \\
2.5842&\ion{Si}{9}&${}^3$P$_{1}$ -- ${}^3$P$_{2}$& \\
3.0285& \ion{Mg}{8}&${}^2$P$_{1/2}$ -- ${}^2$P$_{3/2}$& \\
3.2067& \ion{Ca}{4}&${}^2$P$_{1/2}$ --  ${}^2$P$_{3/2}$& \\
3.3943&\ion{Ni}{3}&${}^1$D$_{2}$ --  ${}^1$D$_{1}$& \\
3.69&\ion{Al}{8}& ${}^3$P$_{1}$ --  ${}^3$P$_{2}$&Uncertain wavelength  \\
3.9342&\ion{Si}{9}&${}^3$P$_{0}$ -- ${}^3$P$_{1}$& \\
4.1594&\ion{Ca}{5}&${}^3$P$_{2}$ -- ${}^3$P$_{1}$&\\ 
4.4867&\ion{Mg}{4}&${}^2$P$_{3/2}$ -- ${}^2$P$_{1/2}$&  \\ 
4.5295&\ion{Ar}{6}&${}^2$P$_{1/2}$ -- ${}^2$P$_{1/2}$&\\
4.6180&\ion{K}{3}&${}^2$P$_{1/2}$ --  ${}^2$P$_{3/2}$&\\
\enddata
\label{tab:coronal_lines}
\end{deluxetable*}

\startlongtable
\begin{deluxetable*}{llll}
\tablecaption{Line identifications for the ejecta}
\tablehead{
\colhead{Wavelength}& \colhead{Ion}& \colhead{Transition}&\colhead{Notes}\\
\colhead{($\mu$m)}&&\\
}
\startdata
1.036&?&&\\
1.062&?&&\\
1.083& \ion{He}{1} & 2s${}^3$S -- 2p${}^3$P  &\\
1.199-1.217&Si I&4s ${}^3$P -- 4p ${}^3$D &\\
1.257&[\ion{Fe}{2}]&   a ${}^6$D$_{9/2}$ -- a ${}^4$D$_{7/2}$&\\
1.283&\ion{H}{1} & 3 -- 5& \\
1.356-1.376& [\ion{Fe}{1}] & a ${}^5$D -- a ${}^5$F  &blend\\
1.444&[\ion{Fe}{1}] &a ${}^5$D$_4$ -- a ${}^5$F$_5$&\\
1.494&[\ion{Fe}{1}] ?&&\\
1.600& [\ion{Fe}{2}]& a${}^4$F$_{7/2}$ -- a${}^4$D$_{3/2}$  & \\
1.607& [\ion{Si}{1}] & ${}^3$P$_{1}$ -- ${}^1$D$_{2}$   & 1.600 -- \\ 
1.644& [\ion{Fe}{2}]&a${}^4$F$_{9/2}$ -- a${}^4$D$_{7/2}$ & 1.664 $\mu$m\\
1.646& [\ion{Si}{1}] &  ${}^3$P$_{2}$ -- ${}^1$D$_{2}$  & blended \\ 
1.664 & [\ion{Fe}{2}]&a${}^4$F$_{5/2}$ -- a${}^4$D$_{1/2}$ & \\ 
1.733&H$_2$& (6,4) O(3)& \\
1.810&[\ion{Fe}{2}]&a ${}^4$F$_{7/2}$ -- a ${}^4$D$_{7/2}$&\\
1.871& \ion{H}{1}  & 3 -- 4 & \\
1.958& H$_2$& (1,0) S(3) & \\
2.059& \ion{He}{1} & 2s${}^1$S -- 2p${}^1$P  & \\
2.122& H$_2$& (1,0) S(1) &  \\
2.166& \ion{H}{1}  & 4 -- 7 & \\
2.248& H$_2$& (2,1) S(1) &  \\
2.407 -- 2.455& H$_2$& (1,0) Q(1-5)  & \\
2.551 -- 2.570& H$_2$& (2,1) Q(1-3) &  \\
2.626& \ion{H}{1}  &  4 -- 6&  \\
2.627& H$_2$& (1,0) O(2) & blend with \ion{H}{1} 2.626 $\mu$m \\ 
2.974&H$_2$&  (2-1)  O(3)&\\
3.846& H$_2$&(0,0) S(13)  &   \\
4.052& \ion{H}{1}  & 4 -- 5 &blend with \ion{He}{1}  4.048 $\mu$m \ \\ %
4.181& H$_2$& 0-0 S(11)  & \\
4.654& \ion{H}{1}  & 5 -- 7 & \\
4.695& H$_2$& (0--0)  S(9)   &   \\
4.926& [\ion{Fe}{1}] & a${}^5$P$_{2}$ -- z${}^7$D$_{3}$ & \\
\enddata
\label{tab:ejectalines}
\end{deluxetable*}

\begin{figure*}[t]
\begin{center}
\includegraphics[width=20cm,angle=90]{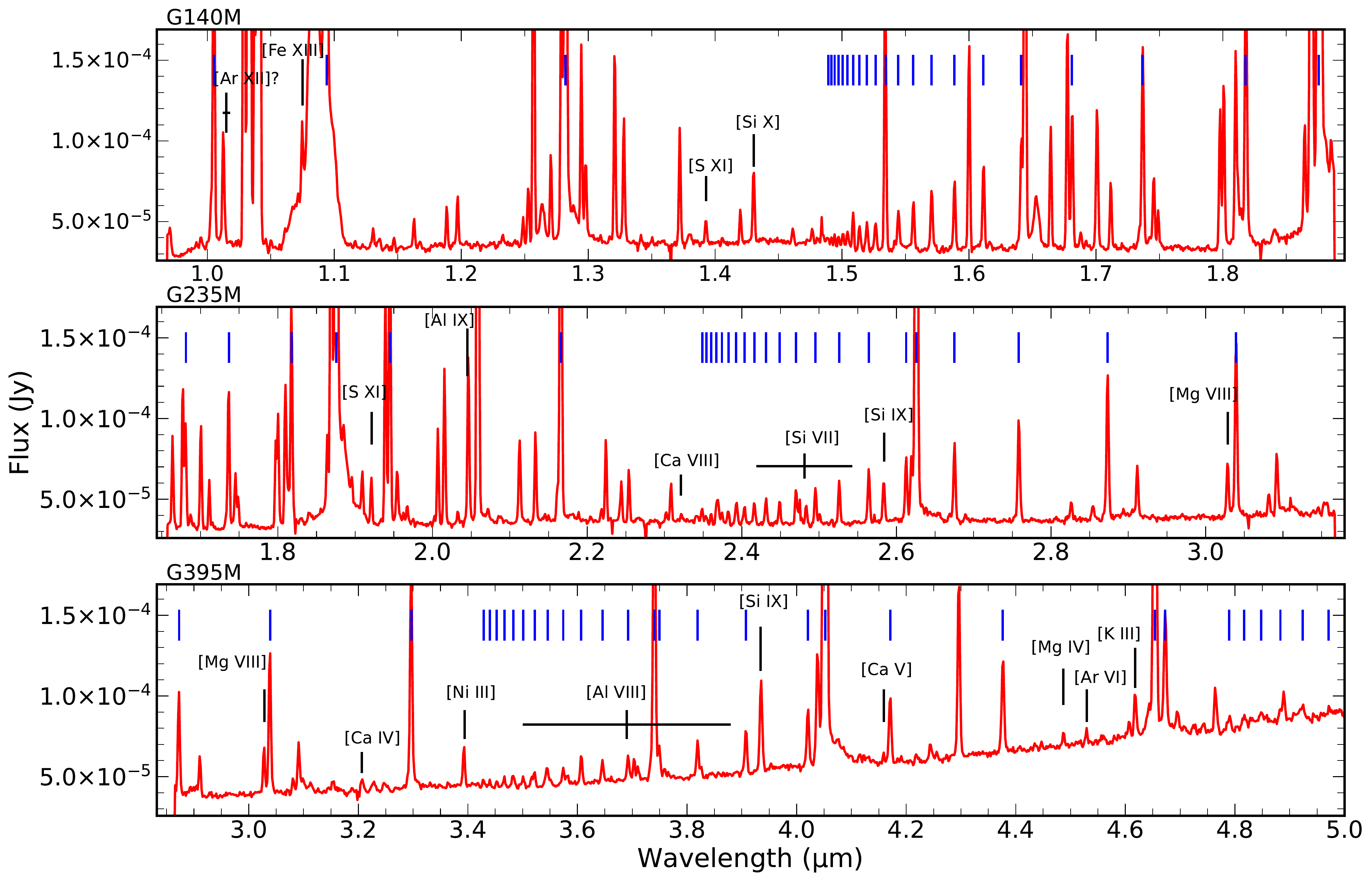}
\caption{Spectrum extracted from 
a 0.13 arcsec$^2$ region in the SW part of the ER with the detected coronal lines marked. We also show the full series of detected lines from the Brackett, Pfund and Humphreys series in blue. 
All coronal line identifications are included in Table~\ref{tab:coronal_lines}. The uncertainties in the laboratory wavelengths are shown as horizontal error bars. To show the faint emission lines, we limit the range of the y-axis compared to Figure \ref{fig:spectra}. The flux scale is here linear while that in Figure \ref{fig:spectra} is logarithmic.}
\label{fig:coronal_spectra}
\end{center}
\end{figure*}

\clearpage

\bibliography{sn87a_refs.bib}{}
\bibliographystyle{aasjournal}

\end{document}